\documentclass[prd,aps,twocolumn,amsmath,nofootinbib,superscriptaddress]{revtex4-2}

\usepackage{amsmath}
\usepackage{amsfonts}
\usepackage{amssymb}	
\usepackage{graphicx}
\usepackage{bm}
\usepackage{color}
\usepackage{commath}
\usepackage{stmaryrd}
\allowdisplaybreaks

\usepackage{hyperref}
\hypersetup{
  colorlinks=true,        % false: boxed links; true: colored links
  linkcolor=blue,         % color of internal links
  citecolor=cyan,         % color of links to bibliography
  urlcolor=cyan            % color of external links
}
\usepackage{natbib,times}

\newcommand{\cf}{cf.~}
\newcommand{\ie}{i.e.,~}
\newcommand{\eg}{e.g.,~}
\begin{document}

\title{Fully general-relativistic simulations of isolated and binary strange quark stars}

\author{Zhenyu Zhu}
\affiliation{Institut f{\"u}r Theoretische Physik,
  Max-von-Laue-Strasse 1, 60438 Frankfurt, Germany}
\affiliation{Department of Astronomy, Xiamen University, Xiamen 361005, China}

\author{Luciano Rezzolla}

\affiliation{Institut f{\"u}r Theoretische Physik, Max-von-Laue-Strasse
  1, 60438 Frankfurt, Germany}
\affiliation{Frankfurt Institute for Advanced Studies,
  Ruth-Moufang-Strasse 1, 60438 Frankfurt, Germany}
\affiliation{School of Mathematics, Trinity College, Dublin 2, Ireland}

\date{\today}

\begin{abstract}
The hypothesis that strange quark matter is the true ground state of
matter has been investigated for almost four decades, but only a few
works have explored the dynamics of binary systems of quark stars. This
is partly due to the numerical challenges that need to be faced when
modelling the large discontinuities at the surface of these stars.
After adopting a suitable definition of the baryonic mass of strange
quark matter (SQM) we introduce a novel technique in which a very thin
crust is added to the equation of state of SQM to produce a smooth and
gradual change of the specific enthalpy across the star and up to its
surface. The introduction of the crust has been carefully tested by
considering the oscillation properties of isolated quark stars, showing
that the response of the simulated quark stars matches accurately the
perturbative predictions. Using this technique, we have carried out the
first fully general-relativistic simulations of the merger of quark-star
binaries finding several important differences between quark-star
binaries and hadronic-star binaries with the same mass and comparable
tidal deformability. In particular, we find that the dynamical mass
loss is $\sim 20\%$ smaller than that coming from a corresponding
hadronic binary. In addition, quark-star binaries have merger and
post-merger frequencies that obey the same quasi-universal relations
derived from hadron stars if expressed in terms of the tidal
deformability, but not when expressed in terms of the average stellar
compactness. Hence, it may be difficult to distinguish the two classes of
stars if no information on the stellar radius is available. Finally,
differences are found in the distributions in velocity and entropy of the
ejected matter, for which quark-stars have much smaller tails. Whether
these differences in the ejected matter will leave an imprint in the
electromagnetic counterpart and nucleosynthetic yields remains unclear,
calling for the construction of an accurate model for the evaporation of
the ejected quarks into nucleons.
\end{abstract}

\maketitle
%------------------------------------------------------------------------
%------------------------------------------------------------------------
\section{Introduction}
\label{sec:introduction}
%------------------------------------------------------------------------
%------------------------------------------------------------------------
The detection of gravitational waves (GWs) from the coalescence of two
compact objects with masses that can be associated to compact stellar
objects has been reported recently by the LIGO-Virgo Scientific
Collaboration~\cite{Abbott2017, Abbott2020, Abbott2020b}. One of these GW
detections, GW170817, was accompanied by an electromagnetic
counterpart~\cite{Abbott2017b, Abbott2017d} and has therefore been
associated with the merger of a binary system of neutron stars which have
long been proposed to be behind the production of short gamma-ray bursts
\cite{Narayan92, Eichler89, Rezzolla:2011, Berger2013b}. Furthermore, the
detection of a kilonova emission through the ejected material and the
subsequent r-process nucleosynthesis \cite{Wanajo2014, Perego2014,
  Just2015, Bovard2017, Thielemann2017b, Radice2018a, Metzger2017,
  Soumi2020, Most2020e} has provided further evidence that the GW signal
in GW170817 must have been produced by the merger of two compact matter
objects.

Combining the knowledge of nuclear physics, general relativity, and
numerical relativity simulations, this detection has certainly helped to
deepen our understanding of cold dense matter equation of state (EOS)
through a tight constraints on maximum mass, radii and tidal
deformability of the neutron stars~\cite{Margalit2017, Bauswein2017b,
  Rezzolla2017, Ruiz2017, Annala2017, Radice2017b, Most2018, Tews2018a,
  De2018, Abbott2018b, Shibata2019, Koeppel2019, Nathanail2021}. Besides
the more conventional scenario of the merger of purely hadronic compact
stars leading to a purely hadronic merged object, other possibilities
have been explored in great detail. One of them is the possibility that
the merging objects were hybrid (or twin) stars~\cite{Fattoyev2017,
  Paschalidis2017, Burgio2018, Montana2018, Gomes2018, Li2018b, Li2020},
that a phase transition to quark matter could have taken place after the
merger~\cite{Most2018b, Bauswein2019, Weih2020}, or that the merger
involved strange-quark stars~\cite{Zhou2017,Drago2018}. All of these
different scenarios are in principle compatible with the GW signal of
GW170817, which was necessarily limited to the inspiral only.

The solution of the equations of general-relativistic hydrodynamic (GRHD)
or general-relativistic magnetohydrodynamic (GRMHD) are indispensable
tools for the accurate modelling of these scenarios and the quantitative
prediction of the signals produced by the inspiral and merger, be it
through the gravitational radiation, the emission of neutrinos, or the
ejection of matter.

Over the years, the scenario of the merger of fully hadronic compact
stars has been studied by GRHD and GRMHD simulations in great
detail~\cite{Shibata99d, Duez2004b, Anderson2007, Baiotti08,
  Rezzolla:2011, Bauswein2011, Bernuzzi2013, Radice2016, Sekiguchi2016,
  Radice2018a, Lehner2016, Ruiz2020b, Bovard2017, Papenfort2018,
  Nedora2020, Most2020e}. Among the numerous results that have been
obtained with these simulations (see \cite{Baiotti2016, Paschalidis2016}
for some reviews), two are particularly relevant for the results
presented in this work. The first one is about the spectral properties of
the GW signal -- both during the inspiral and after the merger -- have
been analysed in great detail and shown to follow quasi-universal
relations in terms of the stellar tidal deformability or
compactness~\cite{Bauswein2012a, Read2013, Bauswein2014, Takami:2014,
  Bernuzzi2014, Takami2015, Rezzolla2016, Bose2017}. The second one is
instead related to the matter ejected at merger and after the merger
\cite{Bauswein2013b, Palenzuela2015, Bovard2017, Dietrich2017c,
  Dietrich2017, Kyutoku2018, Papenfort2018, Radice2018a, Sekiguchi2016,
  Nedora2020, Chaurasia2020}, and the impact that it has on r-process
nucleosynthesis, on the lifetime of the merged object \cite{Gill2019},
and on the maximum mass of compact stars \cite{Margalit2017,
  Rezzolla2017, Nathanail2021}.

While the fully hadronic scenario has been covered in great detail, the
alternative scenarios involving matter in different states -- either
before or after the merger -- has so far been considered less
extensively. The few investigations performed so far have in fact been
limited to the analysis of the post-merger GW signal when a phase
transition sets in after the merger and leads to clear signatures in the
GW signal \cite{Most2018b, Bauswein2019, Weih2020}. In particular, these
studies have highlighted that differences, sometimes significant, can be
found in the post-merge GW spectrum and that the universal relations
found in the case of hadronic stars are obviously broken if a
considerable quark core is produced.

The literature is even scarcer when it comes to simulations exploring the
inspiral and merger of quark-star binaries that composed of pure strange
quark matter (SQM). Indeed, the first and only works so far are more than
a decade old and have been obtained using smooth particle hydrodynamics
and a conformally flat approximation to general
relativity~\cite{Bauswein2009, Bauswein2010}. This is in great part due
to the considerable additional difficulties that the numerical simulation
of these objects implies and that originates from the very sharp decrease
in density and enthalpy at the surface of the quark star.

We recall that the SQM hypothesis was firstly proposed by Witten and
suggesting that SQM, rather than nuclear matter, is the absolute ground
state of matter~\cite{Witten84}. In this hypothesis, SQM is mainly
composed by up, down, and strange quarks, with a small fraction of
electrons also present. Because of the self-bound properties of SQM,
objects composed of this matter can exist in any size, from scales as
small as those of nuclei, to scales as large as that of a compact
star. Indeed, in this hypothesis, a quark star composed of SQM has been
proposed as a possible candidate of compact stars. The properties of
single quark stars have been studied by numerous works \cite{Alcock86,
  Haensel1986, Itoh70, Xu2012, Drago2007, Yu2011, Drago2014a, Drago2016,
  Drago2016c, Drago2016d, Panotopoulos2017, Wiktorowicz2017, Bora2020},
while the study of the binary quark-star mergers have been explored far
less~\cite{Bauswein2009, Bauswein2010, Lai2017b, Drago2018,
  Bucciantini2019, Lai2020}.

The detection of the kilonova signal AT2017gfo associated with the
GW170817 event has been considered by some as a strong evidence against
the existence of SQM. This is because the ejected material of a binary
quark-star merger would be composed by SQM that -- when assumed to be the
most stable form of matter -- cannot be an efficient source of
$r$-process nucleosynthesis. At the same time, recent studies have
highlighted that the SQM scenario can be still conciliated with the
signal from AT2017gfo if the SQM can evaporate into nucleons as a result
of the high temperatures reached after the merger. In this case, most of
ejected SQM from the quark-star binary would have evaporated into
nucleons and could have therefore contributed to the kilonova
signal in AT2017gfo ~\cite{Bucciantini2019, DePietri2019, Horvath2019}.

With the goal of studying the scenario of the merger of quark stars --
and thus explore the possibility of SQM evaporation in far greater detail
-- we have carried out the first fully general-relativistic simulations
of the inspiral and merger of quark stars. Contrasting their evolution
with a system of compact stars that have very similar properties in mass
and compactness but are fully hadronic, we have been able to isolate
three important features of the merger of quarks stars. First, their GW
spectral properties are in agreement with the quasi-universal behaviour
found for hadronic stars during the inspiral, but differs in the
post-merger phase. Second, because of the intrinsic self-boundness of
these objects, the amount of ejected mass is smaller, with binary quark
stars ejecting $\sim 20\%$ less mass than a corresponding hadronic
binary. Finally, as natural to be expected for matter that is colder and
more self-bound, the ejected matter contains considerably smaller tails
in the corresponding distributions of velocity and entropy.

The structure of the paper is as follows. In Sec.~\ref{sec:setup} we
discuss in detail the mathematical and physical setup that was necessary
to develop in order to carry out the numerical simulations. Section
Sec.~\ref{sec:oscil} is instead dedicated to a careful test of the
validity of our approach when considering the oscillation properties of
isolated quark and hadronic stars, and their match with perturbative
studies. Section ~\ref{sec:BQS} is used to present in detail the results
of our simulations, including both the overall dynamics of the merger and
the outcomes in terms of GW signal and ejected matter. Finally, our
conclusions and prospects for future work are presented in
Sec.~\ref{sec:conclusion}, while Appendix \ref{sec:baryon mass} provides
details on the technique that can be used when considering different
values of the baryon mass.

%------------------------------------------------------------------------
%------------------------------------------------------------------------
\section{Mathematical and Numerical Setup}
\label{sec:setup}
%------------------------------------------------------------------------
%------------------------------------------------------------------------

%------------------------------------------------------------------------
\subsection{Equations of state}
\label{sec:EOS}
%------------------------------------------------------------------------

The EOS of the SQM employed in our simulations was chosen to be the
MIT2cfl EOS \cite{Zhou2017, zhu2020}. This EOS makes use of the MIT bag
model with additional perturbative QCD corrections~\cite{Fraga2001,
  Alford2005, Li2017, Zhou2017}, and satisfies the constraints of having
a maximum mass above two solar masses as required by the observations
\cite{Demorest2010, Antoniadis2013}, and a tidal deformability compatible
with the constraints from GW170817. In this EOS a colour-flavor-locked
(CFL) phase is assumed to be present and consequently the number
densities of all the flavor of quarks [up (u), down (d), and strange (s)
  quarks] are the same, \ie
\begin{equation}
  n_{u} = n_{d} = n_{s}\,.
  \label{eq:cfl}
\end{equation}
As a result, the baryon number density defined as
\begin{equation}
n_{_{\rm B}} :=
  \frac{1}{3}(n_{u} + n_{d} + n_{s})\,,
\end{equation}
can be directly converted to the rest-mass density $\rho$ that enters in
the hydrodynamic equations and is evolved numerically, as $\rho :=
m_{_{\rm B}} n_{_{\rm B}}$, where $m_{_{\rm B}}$ is the average
baryonic mass. We note that while the value of $m_{_{\rm B}}$ is well
defined for a baryon, this is not the case when considering SQM. In
particular, we recall that the definition of the baryonic mass for
hadronic matter is given by the value of total energy density per
baryon at vanishing pressure, \ie
\begin{eqnarray}
  m_{_{\rm B}} := \frac{e(p=0)}{n_{_{\rm B}}(p=0)}\,.
  \label{eq:bmass}
\end{eqnarray}
As a result, when considering the MIT2cfl EOS employed here, we obtain
that the baryon mass is $m_{_{\rm B}}\simeq 850\,{\rm MeV}$. We note that
this value is smaller than the one assumed for hadronic matter, \ie
$940\,{\rm MeV}$ (see, \eg \cite{DePietri2019}), but has been employed
also by other groups (see, \eg \cite{Bauswein2009, Bhattacharyya2016,
  Drago2020}). Two comments should be made at this point. First, the
definition Eq.~(\ref{eq:bmass}) naturally implies that at zero pressure
the specific internal energy is also zero, \ie $\epsilon(p=0)=0$, so that
the specific enthalpy at the stellar surface is $h(p=0)= 1 + \epsilon +
p/\rho=1$, as one would expect. Second, as we will comment in more detail
in Appendix \ref{sec:baryon mass}, a different value of the baryon mass
inevitably introduces a discontinuity at the stellar surface, thus
calling for a suitable rescaling in order to carry out the numerical
simulations.

Possibly the most serious challenge in modelling compact stars made of
SQM is that -- because of their self-bound property -- the surface of
such objects is characterized by a sharp transition at the stellar
surface, where the pressure goes to zero at a nonzero rest-mass
density. As a result, a large, intrinsic density jump is present at the
stellar surface; by contrast, hadronic stars have surfaces near which the
density rapidly decreases, but that goes to zero when the pressure is
zero. When considering static solutions, as for example when constructing
stellar models of isolated quark stars, such a density jump can be
handled analytically by matching the stellar interior with the exterior
vacuum ~\cite{Damour:2009, Postnikov2010, Zhou2017, zhu2020}. However, in
the context of a hydrodynamic simulation, such a discontinuity represents
the exemplary condition for the development of a strong shock that would
lead to an artificial oscillation in the best-case scenario or to a
numerical failure in the most realistic case. Clearly, a treatment aimed
at smoothing this strong discontinuity into a region with small but
finite size is necessary for a numerical evolution.

A simple solution at the level of the EOS may consist in the
introduction of a polytropic piece in the pressure dependence from the
rest-mass density, \ie $p=k\rho^{\Gamma}$ with $k=8.12$ and
$\Gamma=1.90$, thus effectively introducing a thin but nonzero ``crust''
in the quark star. The presence of a thin crust in a quark star and its
implications have been studied in detail in a number of
works~\cite{Kettner95, Huang97, Wu2020}. The introduction of a 
crust changes, at least in principle, the tidal deformability. In 
practice, however, the change is extremely small and of $1.3\%$ only. 
More precisely, the tidal deformabilities for a quark star with and 
without crust are 789.3 and 778.9, respectively.

A few important aspects of our novel approach need to be remarked at this
point. First, although artificial, the introduction of the thin crust does not
provide a perceptible variation to the global properties. While we will
demonstrate this in Sec. \ref{sec:oscil}, where we will compare the
oscillation properties of an isolated star with the perturbative
expectations, it suffices to say here that the variation on the
gravitational mass after the introduction of thin crust is minute, \ie
$\sim 5\times 10^{-3}\,M_\odot$, as is its spatial extension, which is
restricted to two grid cells and therefore has a width of $\simeq
240\,{\rm m}$. 

  Second, since we have introduced a thin crust, our compact star
  could be assimilated to a ``hybrid star'' as it is effectively composed
  of two regions: a quark-matter and a nuclear-matter region, although
  the latter is extremely thin. Indeed, simulations of this type of
  compact stars were performed and investigated in great detail in
  Refs.~\cite{Tsokaros2019b, Tsokaros2020b}. However, this similarity is
  potentially misleading because a strange quark star is actually
  composed of SQM that represents the ground state at any density. On the
  contrary, the quark matter in a hybrid star could exist as the ground
  state only if the density is sufficiently high, namely, at least larger
  than the saturation density.  Hence, we prefer to regard ours as a
  strange star with a thin crust rather than a hybrid star.

Finally, the handling of the MIT2cfl EOS for the actual merger requires
the addition of a thermal part to the EOS. We do this in close analogy
with what done routinely for simulations of hadronic stars described by
cold EOSs. In essence, we account for the additional shock heating during
the merger and post-merger phases by including thermal effects via a
``hybrid EOS'', that is, by adding an ideal-fluid thermal component to
the cold (subscript ``c'' below) EOS \cite{Rezzolla_book:2013}
\begin{eqnarray}
p & = & p_{\rm c} + p_{\rm th}\,, 
\label{eq:therm1}  \\
p_{\rm th} & = & \rho\epsilon_{\rm th} (\Gamma_{\rm th} - 1)\,, 
\label{eq:therm2}  \\
\epsilon_{\rm th} & = & \epsilon - \epsilon_{\rm c}(\rho)\,,
\label{eq:therm3}
\end{eqnarray}
where $\Gamma_{\rm th}=1.75$ is the thermal adiabatic index.

Since we find it important to contrast the dynamical behaviour of merging
quark stars with the corresponding one of hadronic stars with the same
total mass, we have also considered the merger of a binary system subject
to a hadronic EOS for comparison. Our choice has fallen on the DD2 EOS
\cite{Typel2010}, which is compatible with the present observational
constraints, and whose tidal deformability for a $M=1.35\,M_\odot$ star
is close to that of a quark star of the same mass. At the same time, the
radius of the quark star is significantly smaller: $R=11.81\,\rm{km}$ for
the MIT2cfl quark star versus $R=13.21\,\rm{km}$ for DD2 hadronic star.

In the framework of a comparative assessment of the dynamics of SQM and
hadronic-matter binaries, and despite the fact that the hadronic DD2 EOS
is a finite-temperature EOS employed in many GRHD or GRMHD
simulations~\cite{Radice2017a, Bovard2017, Radice2018a, hanauske2019a,
  Most2019b}, we have adopted a hybrid-EOS approach [\cf
  Eqs.~(\ref{eq:therm1})--(\ref{eq:therm3})] also for the DD2 EOS, of
which we have retained only the zero-temperature slice. An immediate
disadvantage of this approach is that the consistent knowledge of some
thermodynamical quantities, such the temperature or the entropy, are
missing and need to be estimated in alternative manners. In this case,
additional assumptions -- some of which are not necessarily realistic --
are required to extract these quantities, at least to some
degree. However, since the same approximations are made for both EOSs,
the expectation is that the systematic differences we can find in this
way will persist also when considering more advanced and
temperature-dependent EOSs for SQM.

More specifically, in the case of ideal-fluid EOS, the temperature $T$ is
proportional to the average kinetic energy per particle and we can therefore
express the specific internal energy $\epsilon$ as
\begin{equation}
\epsilon = \frac{k_t T}{m_{_{\rm B}}}\,, 
\end{equation}
where $k_t$ is a constant and $m_{_{\rm B}}$ is mass per baryon. We
extend this expression to our EOSs by rewriting it as
\begin{eqnarray}
  \epsilon = k_t \frac{(T - T_{\rm c})}{m_{_{\rm B}}} + \epsilon_{\rm c}(\rho)\,,
  \label{eq:eps_t}
\end{eqnarray}
where $T_{\rm c}$ is the temperature of the cold part of the EOS and
which we take to be to $T_{\rm c}=0.01\,{\rm MeV}$. Using now
Eq.~(\ref{eq:eps_t}) and recalling that for transformations at constant
density $d\epsilon = Tds$, we can compute the specific entropy as
\begin{eqnarray}
  s = \frac{k_t}{m_{_{\rm B}}}\log\left(\frac{\epsilon - \epsilon_{\rm
      c}}{k_t T_{\rm c}/m_{_{\rm B}}} + 1\right)\,.
  \label{eq:entropy}
\end{eqnarray}
Finally, specifying $k_t = 20$ for both EOSs, we can compute the entropy
per baryon $s_{_{\rm B}} = m_{_{\rm B}} s$ once $\epsilon$ and
$\epsilon_{\rm c}$ are known.

%------------------------------------------------------------------------
\subsection{Numerical setup: Initial data}
%------------------------------------------------------------------------

The initial data for binary stars was generated making use of the
publicly available \texttt{Lorene}
code~\cite{Gourgoulhon-etal-2000:2ns-initial-data}, which is a
multi-domain spectral-method code computing quasi-equilibrium
irrotational binary configurations of compact stars. Using this code, and
as discussed above, we have computed initial binary configurations of
both quark stars and of hadronic stars. Independently of whether we have
considered the $\rm{MIT2cfl}$ or the $\rm{DD2}$ EOS, the properties of
the binary have been set to be the same: the initial separation in the
binary was set to be $45\,{\rm km}$ and the two stars have the same mass
of $M=1.35\,M_\odot$.

It is useful to remark that the introduction of a thin crust for the
quark star was important also for the calculation of the initial
data. Furthermore, the computation of the solution of binary quark stars
with \texttt{Lorene} required extra care. In particular, because of the
steep drop in rest-mass density across the crust of the quark star and of
the presence of discontinuities in higher-order derivatives at the
crust-core interface (we recall that the EOS is continuous but with
discontinuous derivatives at crust-core interface), a single interior
coordinate domain employed to cover the quark star turned out to be
insufficient and prevented the convergence to an accurate solution.
Fortunately, however, the addition of a second domain at higher
resolution to cover the crust was sufficient to yield the convergence to
an accurate solution.

\begin{table}[t]
  \renewcommand{\arraystretch}{1.3}
  \caption{Properties of the quark and hadron stars considered here, and
    distinguished in whether they refer to the isolated configurations or
    to binaries. Reported are the mass $M$, the baryon mass $M_b$, the
    radii $R$, and the tidal deformability $\Lambda$. In addition, in the
    case of the binaries, we also report the relevant frequencies of the
    gravitational-wave signal ($f_{\rm max}, f_1$, and $f_2$), the
    ejected matter $M_{\rm ej}$ and the ejected baryon number 
    $N_{\rm B}^{\rm ej}$.}
\begin{center}
\begin{tabular}{|l|c|r|r|} 
  \hline
  \hline
     & & Quark star & Hadronic star \\ 
   \hline
${\rm EOS}$          & ${\rm type}$       & ${\rm MIT2cfl}$  & ${\rm DD2}$   \\ 
\hline
$M\ [M_\odot]$ & $({\rm isolated})$        & $1.40$  & $1.40$     \\
$M_{b}\ [M_\odot]$ & $({\rm isolated})$    & $1.57$  & $1.53$     \\
$R\ [{\rm km}]$ & $({\rm isolated})$      & $11.90$  & $13.22$    \\
$\Lambda$ & $({\rm isolated})$    	      & $657.78$ & $698.72$   \\
\hline
$M\ [M_\odot]$ & $({\rm binary})$          & $1.35$  & $1.35$       \\
$M_{b}\ [M_\odot]$ & $({\rm binary})$      & $1.50$  & $1.47$       \\
$R\ [{\rm km}]$ & $({\rm binary})$       & $11.81$  & $13.21$       \\
$\Lambda$ & $({\rm binary})$	     & $789.32$ & $857.69$      \\
$f_{\rm max}\ [{\rm kHz}]$   & $({\rm binary})$	     & $1.684$  & $1.644$       \\
$f_1\ [{\rm kHz}]$ & $({\rm binary})$          	     & $1.919$  & $1.845$       \\
$f_2\ [{\rm kHz}]$ & $({\rm binary})$                   & $2.399$  & $2.436$       \\
$M_{\rm ej}\ [10^{-3}\,M_\odot]$ & $({\rm binary})$        & $2.68$ & $2.94$          \\
$N_{\rm B}^{\rm ej}\ [10^{54}]$ & $({\rm binary})$ & $3.520$ & $3.533$ \\
  \hline
  \hline
\end{tabular}
\label{tab:gw}
\end{center}
\end{table}

\begin{figure*}[t]
  \centering
  \includegraphics[width=0.49\textwidth]{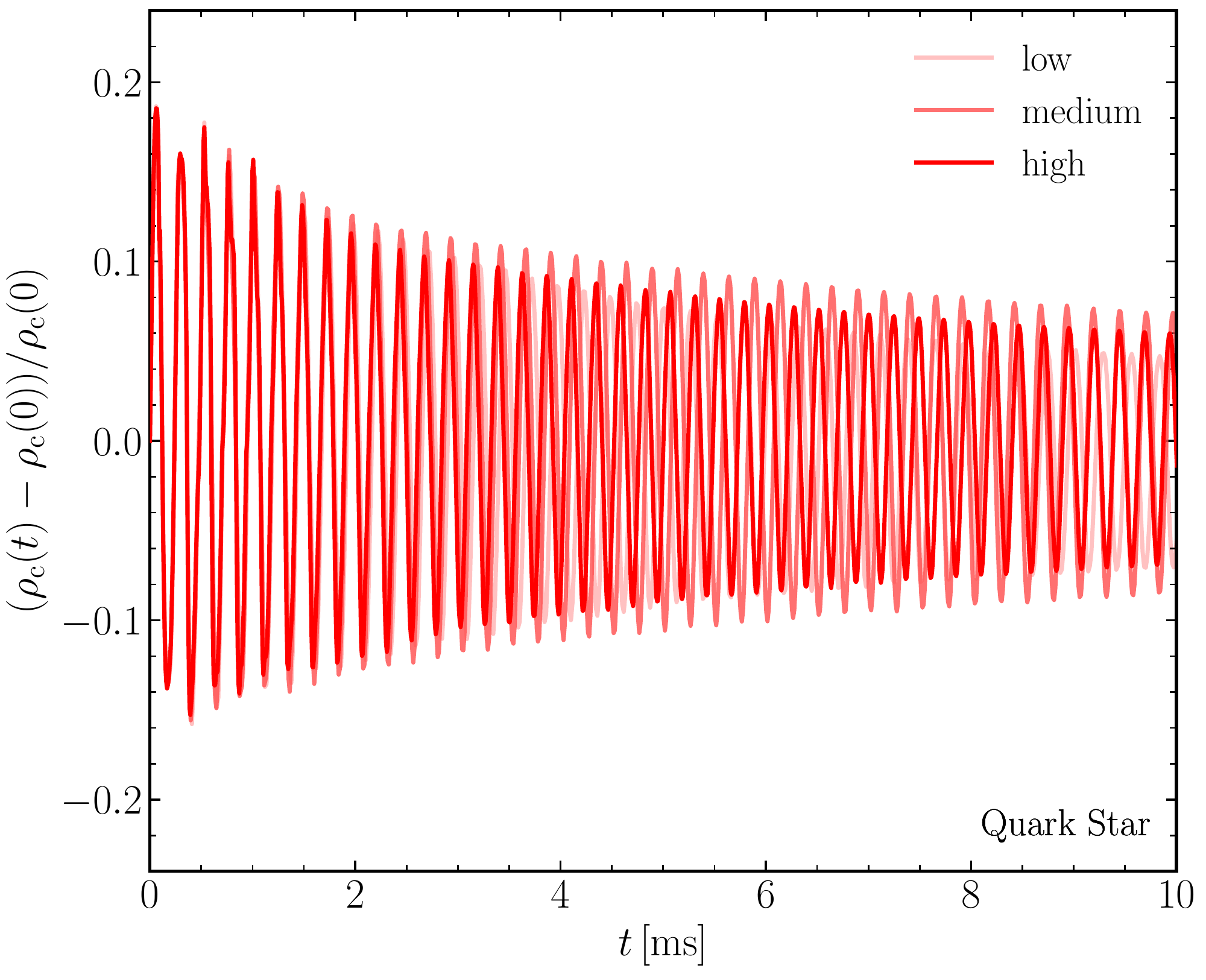}
  \includegraphics[width=0.49\textwidth]{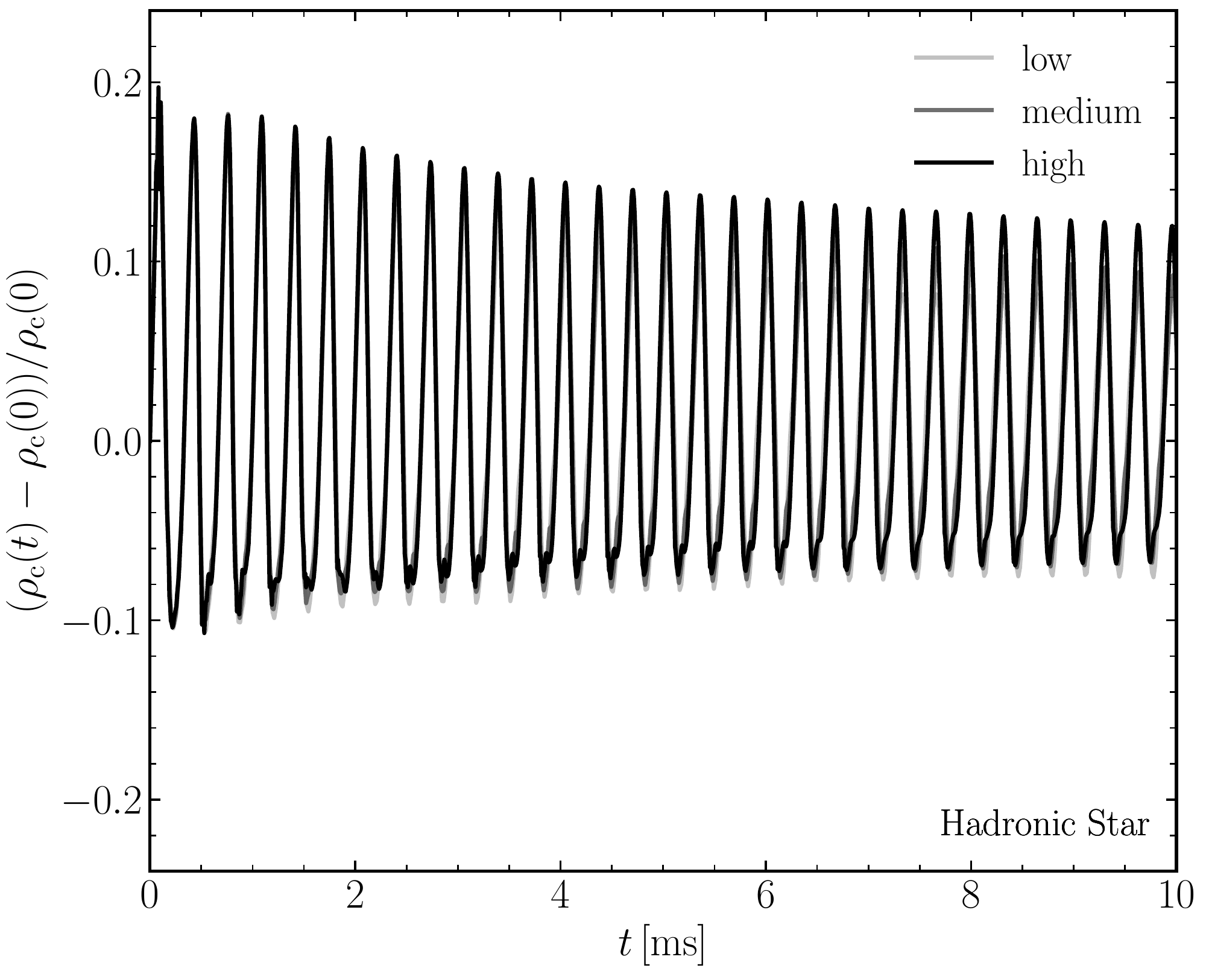}
  \caption{Evolutions of the normalised variations in the stellar central
    rest-mass density $\rho_c(t)$ for either an oscillating quark star
    (left panel) or for a hadronic star (right panel). Curves of
    different shading refer to different resolutions, with the
    high-resolution data being shown with the darkest shade.}
  \label{fig:oscil}
\end{figure*}

%------------------------------------------------------------------------
\subsection{Numerical setup: evolution equations}
%------------------------------------------------------------------------

The simulations presented here have been performed with the publicly
available GRHD code \texttt{WhiskyTHC}~\cite{Radice2012a, Radice2013b,
  Radice2013c}, which is a high-order, fully general-relativistic code
for the solution of the equations of relativistic hydrodynamics and
compatible with the \texttt{Einstein Toolkit}
\cite{EinsteinToolkit_etal:2020_11}. The hydrodynamic equations were
solved by using a high-resolution shock-capture
(HRSC)~\cite{Rezzolla_book:2013} approach having with Local
Lax-Friedrichs (LLF) flux-split and Monotonicity Preserving 5th-order
(MP5) reconstruction method~\cite{suresh_1997_amp, mignone_2010_hoc}
under the finite-difference scheme. The spacetime was instead evolved by
implementing the CCZ4 formulation~\cite{Alic:2011a, Alic2013} through the
finite-differencing code \texttt{McLachlan} \cite{Brown:2008sb}.

In order to cover a large enough spatial domain and hence compute
accurately the information on the ejected matter while keeping a high
resolution on the compact stars, an adaptive-mesh refinement (AMR)
approach was employed and handled by the \texttt{CARPET}
driver~\cite{Schnetter:2003rb}. The number of refinement levels depends
on the problem considered and was of $3$ levels in the case of isolated
stars and of $6$ levels in the case of merging binaries. In this way, the
corresponding total extents of the computational domain were set to $24\,
M_\odot\ (36\, {\rm km})$ and $2048\, M_\odot\ ( 3051\, {\rm km})$,
respectively. Furthermore, each of the scenario considered -- isolated
stars or binary mergers -- has been evolved with simulations at different
resolutions. More specifically, for the finest grid we have used three
resolutions in the case of isolated stars, \ie $dx = 0.16\,
M_\odot\ (236\, {\rm m})$, $dx = 0.12\, M_\odot\, (177\, {\rm m})$, and
$dx = 0.08\, M_\odot\ (118\, {\rm m})$, which we will indicate as
``low'', ``medium'', and ``high'' resolution in following. Similarly, for
both quark and hadronic stars, we have employed two resolutions in the
case of merging binaries, \ie $dx = 0.25\, M_\odot\ (369\, {\rm m})$ and
$dx = 0.16\, M_\odot\ (236\, {\rm m})$, which we will indicate as ``low''
and ``high'', respectively. Note that a resolution of $dx = 0.16\,
M_\odot\ (236\, {\rm m})$, and even lower ones, (see \eg
\cite{DePietri2016,Depietri2018}) have been used routinely in simulations
of binary neutron stars and has been shown to be sufficient to capture
accurately the dynamics of the system.

%------------------------------------------------------------------------
%------------------------------------------------------------------------
\section{Results: isolated quark stars}
\label{sec:oscil}
%------------------------------------------------------------------------
%------------------------------------------------------------------------

As a first but very important test of the correct implementation of the
rescaling procedure presented in the previous Section, we next discuss
the results from the evolution of isolated and oscillating quark stars,
comparing with the corresponding evolutions obtained with fully hadronic
stars. The analytical formulation and the numerical computation of the
spectral properties of oscillating compact stars is a very old one and
has been studied in detail in the literature~\cite{Chandrasekhar64,
  Campolattaro1, McDermott1988, Rezzolla2002b}, also in the case of quark
stars \cite{Panotopoulos2017, Bora2020}. Similarly, the investigation of
the dynamical response of a relativistic star to a perturbation has been
studied extensively in the literature (see, \eg \cite{Font99, Font02c,
  Baiotti04} for some of the initial works).

\begin{figure*}[t]
  \centering
  \includegraphics[width=0.49\textwidth]{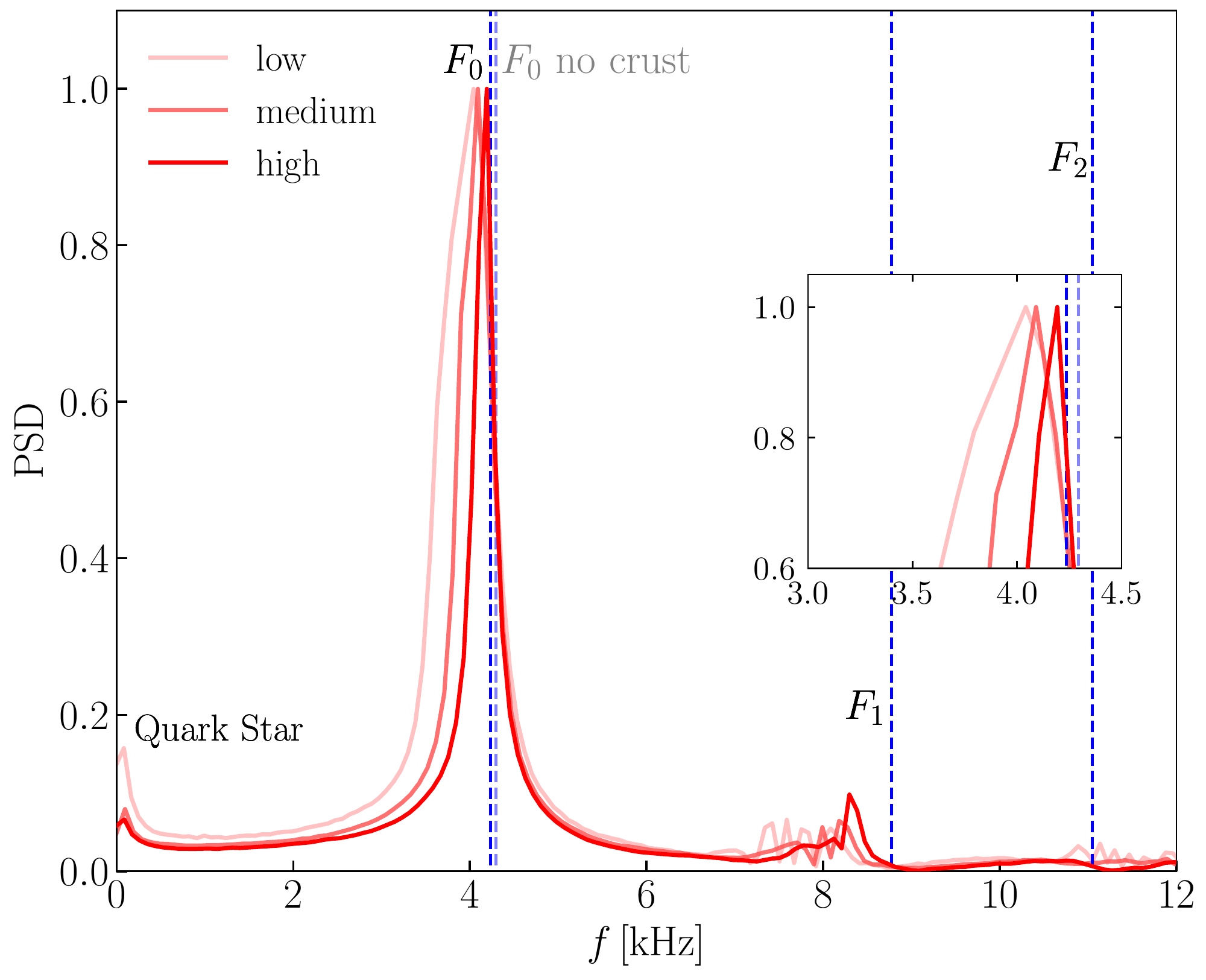}
  \includegraphics[width=0.49\textwidth]{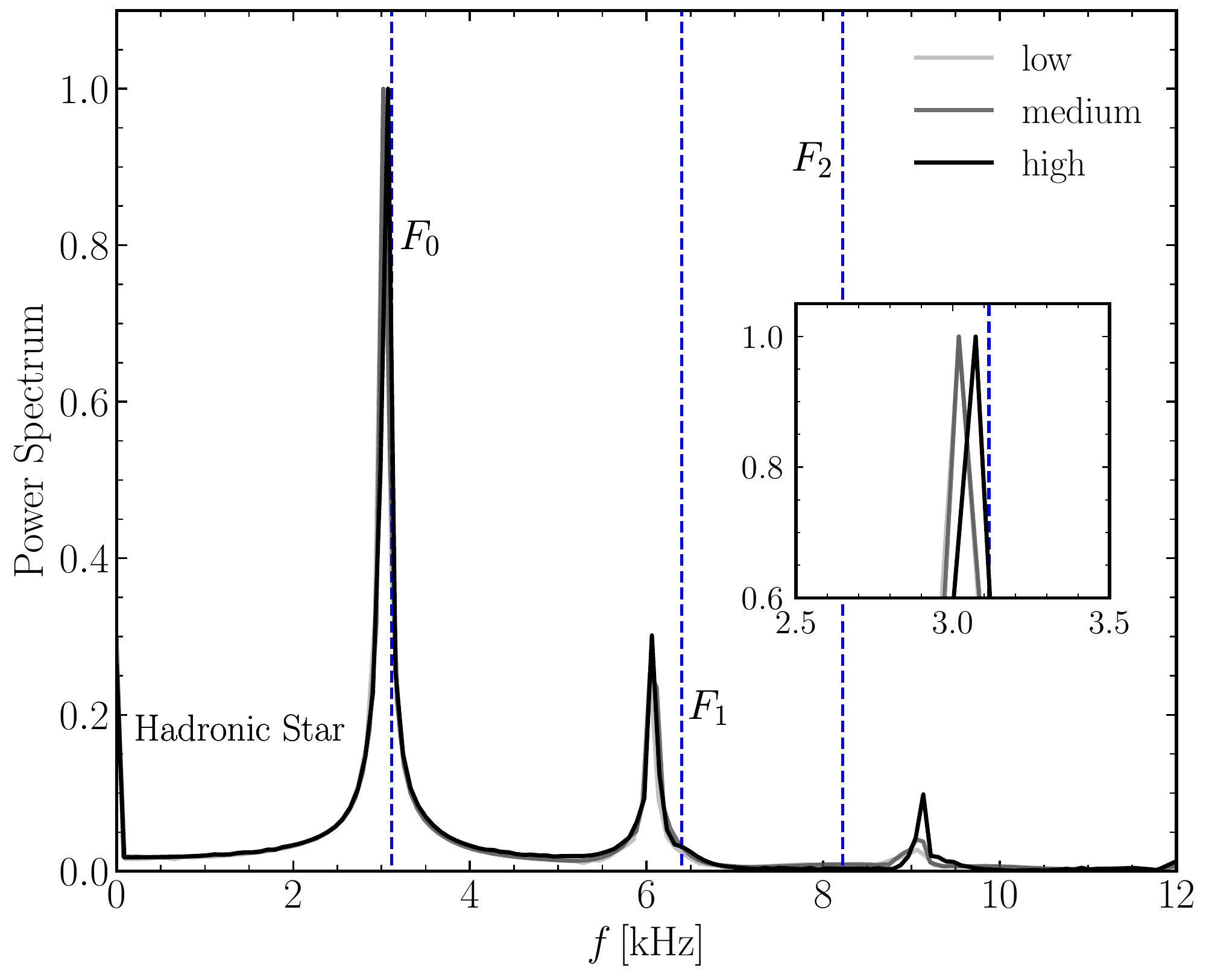}
  \caption{Power spectral densities of the timeseries in
    Fig. \eqref{fig:oscil} when compared with the perturbative
    predictions for the eigenfrequencies, reported as vertical blue
    dashed lines. The left panel refers to a quark star, while the right
    one to a hadronic star; the same convention is followed for the
    different types of lines. In both cases, the insets show a
    magnification near the frequency of the $F_0$ mode. Note that the
    accuracy of the match increases as the resolution is increased.}
  \label{fig:psd_oscil}
\end{figure*}

Because this serves only as a test and not to study the spectral
properties of such stars, the oscillations we have considered are simple
quasi-radial oscillations of initially static stars, whose perturbative
eigenfrequencies of can be obtained easily by solving two ordinary
differential equations [see Eqs. (11)--(12) in Ref.~\cite{Bora2020}] that
are coupled with the equations of stellar equilibrium, \ie the
Tolmann-Oppenheimer-Volkoff (TOV) equations (see, \eg
\cite{Rezzolla_book:2013}). To make the comparison meaningful, both the
isolated MIT2cfl quark stars and the hadronic DD2 stars have the same
mass of $M=1.4\,M_{\odot}$. Solving the perturbation equations together
with the TOV equations, we obtain the eigenfrequencies $F_i$, for
different mode numbers $i$. In practice, we limit ourselves to the
fundamental model $F_0$ and the first overtone $F_1$. Note that both the
static models and the eigenfrequencies are computed after introducing the
crustal prescription in the EOS, as described in the previous
Section. For the dynamical evolutions, on the other hand, we introduce an
initial purely radial perturbation in the radial velocity with an
eigenfunction corresponding to an $n=1$, $\ell=0$ oscillation mode under
the Cowling approximation (\ie with a spacetime being held fixed)
\cite{Cowling41}, and an amplitude of $3\%$.

As customary, we report in Fig.~\ref{fig:oscil} the normalised variation
in the stellar central density $\rho_c(t)$ for either the quark star as
discussed above (\ie with a thin crust, left panel) or for the hadronic
star (right panel). Curves of different shading refer to different
resolutions, with the high-resolution data being shown with the darkest
shade. Note that, especially for the quark star, the behaviour of the
oscillations shows aspects of strong nonlinearity, \ie deviation from a
simple harmonic oscillation, during the first few millisecond, but these
then disappear rather soon and the oscillation becomes rather regular. As
to be expected, these nonlinear features becomes more marked as the
resolution is increased and the deviations from a perfect conservation of
rest-mass are $\lesssim 10^{-5}\%$.

Note also the imprint of different resolutions is more evident in the
quark star than in the hadronic one. A phase difference can in fact be
measured in the low- and medium-resolution evolutions for the quark star,
which is instead absent for the hadronic star\footnote{Note that the
  amplitude of the oscillations is not necessarily related to the
  resolution employed in the simulation. Indeed, the amplitude evolution
  depends on a number of factors, such as the ability to capture the
  eigenmodes of the oscillations and the occurrence of small shocks at
  the interface between the star in the fluid outside the star. As a
  result, oscillations of different amplitudes can be obtained even when
  using the same resolution but different reconstruction scheme (see the
  detailed discussion in Ref. \cite{Radice2012a}.}. This is again easy
to interpret as due to the much sharper density jump at the stellar
surface in the case of a quark star, which requires therefore a higher
resolution to produce a consistent behaviour. However, for both stars,
the low resolution is sufficient to capture a convergent behaviour and
the accuracy simply increases as the resolution is increased. Finally, it
should be noted that while the quark star exhibits oscillations that are
almost symmetric with respect to the unperturbed value, the hadronic one
does not. This is due to the fact that for the latter, the oscillations
are accompanied by a significant contribution from higher-order modes
(see discussion below).

\begin{figure*}[t]
  \centering
  \includegraphics[width=0.85\textwidth]{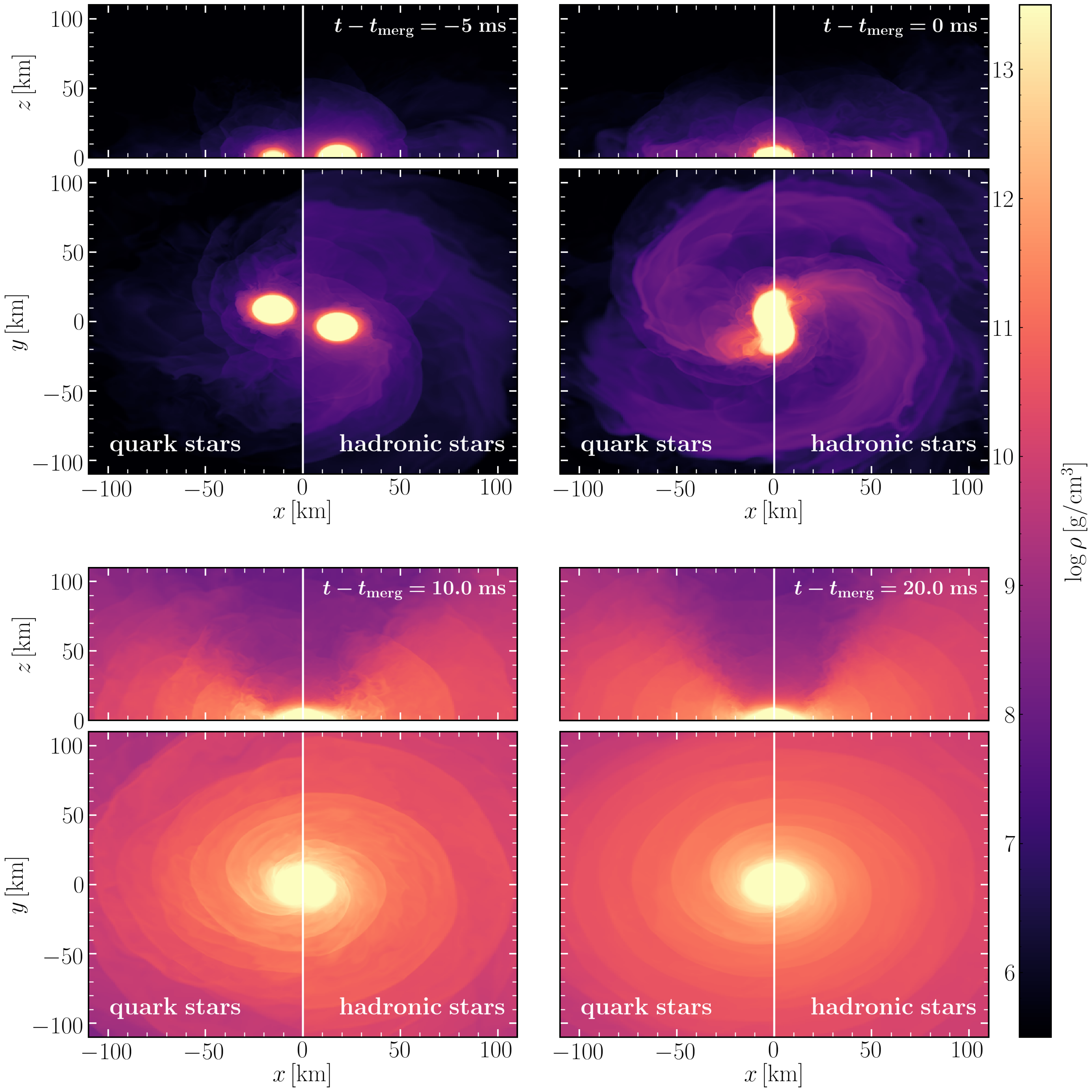}
  \caption{Small-scale (\ie $110\,{\rm km}$) views of the rest-mass
    density at different but characteristic times during the merger of
    the two classes of binaries [\ie $5\,{\rm ms}$ before merger
      (upper-left), the merger time (upper-right), $10\,{\rm ms}$ (lower
      left), and $20\,{\rm ms}$ (lower right) after merger]. Each panel
    offers views of the $(x,y)$ plane (bottom parts) and of the $(x,z)$
    plane (top parts), respectively. Note that, for each panel, the left
    part refers to quark stars, while the right one to the hadronic
    stars.}
  \label{fig:rho_2D_zoom}
\end{figure*}
\begin{figure*}[t]
  \centering
  \includegraphics[width=0.85\textwidth]{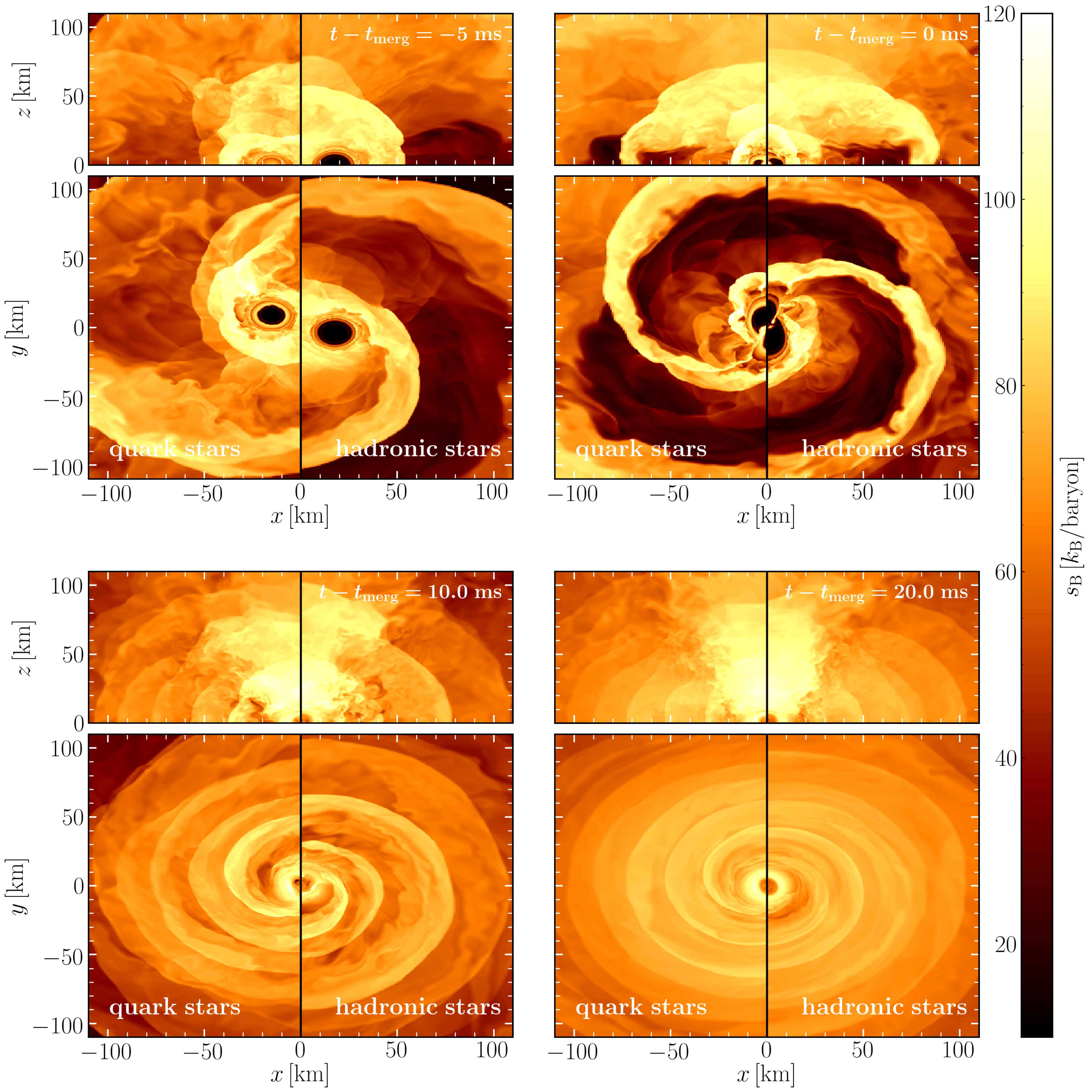}
  \caption{Same as in Fig.~\ref{fig:rho_2D_zoom}, but for the entropy per
    baryon.}
  \label{fig:s_2D_zoom}
\end{figure*}

Figure \ref{fig:psd_oscil} helps to compare and contrast the evolutions
of quark and hadron stars by taking the Fourier transforms of the
timeseries in Fig. \eqref{fig:oscil} and comparing the resulting power
spectral densities (PSDs) with the perturbative prediction for the
eigenfrequencies, reported as vertical blue dashed lines. In this way, it
is apparent that the match between the dynamical and the perturbative
results is worse for the quark star, especially at low resolutions (see
the insets). However, as the resolution is increased, both evolutions
nicely converge to the perturbative values and differ from them with
comparable errors, namely, $1.1\%$ ($3.1\%$) and $1.3\%$ ($3.5\%$) for
the quark and hadron stars at high (medium) resolution,
respectively. Note also that the low-resolution of the quark star has a
considerable amount of noise near the position of the $F_1$ frequency,
but that this noise decreases with increasing resolution, leading to the
clear appearance of an overtone in the high-resolution run. Finally, note
that the amount of power in the two overtones is much larger in the case
of the hadronic star, which exhibits a clear second overtone. This
behaviour in indeed visible already in the timeseries reported in
Fig.~\ref{fig:oscil} (right panel), which exhibits rapid variations near
the maxima and minima of the oscillations. Also reported for a
comparison and marked with a transparent blue dashed line is the $F_0$
eigenfrequency of a strange star without a crust; note that this is
larger and that the simulation do not converge to this value.

In summary, the results presented here on the dynamics of isolated quark
stars show that the approach presented in Sec. \ref{sec:EOS} to handle
the stellar surface is not only properly implemented, but that it also
leads to evolutions that are able to reproduce the spectral properties of
quark stars and yield numerically consistent evolutions. At the same
time, they show that, in general, quark stars require higher resolutions
than the ones that would be needed for a hadronic star with comparable
properties in terms of mass and tidal deformability. Once again, this is
the inevitable drawback of having to model stars with a strong
self-confinement and sharper density profiles at the surface.

%------------------------------------------------------------------------
%------------------------------------------------------------------------
\section{Results: binary quark stars}
\label{sec:BQS}
%------------------------------------------------------------------------
%------------------------------------------------------------------------

In this section we discuss the dynamics of a binary quark-star merger and
contrast it with the corresponding merger obtained with a hadronic
EOS. Before entering in the details, and notwithstanding that these are
the first fully general-relativistic simulations of the merger of SQM
compact objects, it is useful to stress what are the limitations of our
approach. First, the treatment of the thermal part of the EOS is
necessarily approximate and modeled with a hybrid EOS approach for both
EOSs [\cf Eqs. \eqref{eq:therm1}--\eqref{eq:therm3}]. As a result of this
(forced) choice, the modelling of neutrinos, \eg via a leakage
\cite{Galeazzi2013}, or more advanced approaches \cite{Radice2016,
  Weih2020b}, is not possible. Second, the modelling of the
quark-evaporation processes in the ejected matter, is de-facto
ignored. While both processes are expected to have a significant impact
on the electromagnetic kilonova signal and on the nucleosynthesis, they
are not expected to play an important role in the gravitational-wave
signal, which can therefore be considered robust. Third, for obvious
computational costs, the two resolutions used here for the binaries are
rather low, and the highest one would correspond to the ``low''
resolution in the case of the isolated stars discussed in the previous
Section. However, as demonstrated above, such a resolution is sufficient
to provide a sufficiently accurate description of the stellar behaviour
given that the relative error in the PSD for the low-resolution quark
star is $\lesssim 4.6\%$. More importantly, by comparing and contrasting
quark and hadronic stars with similar properties we can easily capture --
even at these low resolutions -- the systematic differences between the
two stellar types, which we believe are therefore robust.

%------------------------------------------------------------------------
%------------------------------------------------------------------------
\subsection{Overview of the dynamics}
\label{ssec:overview}
%------------------------------------------------------------------------
%------------------------------------------------------------------------

The main features of the dynamics of the binary systems of quark and
hadronic stars are rather similar. Before the merger, the stars in the
binary systems inspiral towards each other as a result of energy and
angular moment loss via gravitational-wave emission. During this stage,
the matter effects are dominated by the tidal deformability of the two
stars. This stage is shown in the top-left panels of
Figs. \ref{fig:rho_2D_zoom}--\ref{fig:s_2D_zoom}, which report the
small-scale (\ie $110\,{\rm km}$) views of the rest-mass density
(Fig. \ref{fig:rho_2D_zoom}) or of the entropy per baryon
(Fig. \ref{fig:s_2D_zoom}). Different panels refer to different but
characteristic times during the merger [\ie $5\,{\rm ms}$ before merger
  (upper-left), the merger time (upper-right), $10\,{\rm ms}$ (lower
  left), and $20\,{\rm ms}$ (lower right) after merger] and offer views
of the $(x,y)$ plane (bottom parts) and of the $(x,z)$ plane (top parts),
respectively. More importantly for our comparison, for each panel, the
left part refers to quark stars, while the right one to the hadronic
stars. While not straightforward to analyse,
Figs. \ref{fig:rho_2D_zoom}--\ref{fig:s_2D_zoom} are rich of information
and help obtain a comprehensive overview of the properties of the
dynamics of the two binaries.

It should be noted that already in the inspiral phase, the quark-star
binary loses a smaller amount of material from the surface. While this
matter is not significant both in terms of gravitational-wave emission
and nucleosynthesis (the material has very small velocities, is
distributed essentially isotropically, and is gravitationally bound in
good part), it already shows a feature we will encounter again below.

As the two stars approach each other, the amplitude of the emitted
gravitational waves increases, achieving a maximum when the stars
merge. At this point in time, intense tidal forces and strong shocks
induce a significant and dynamical ejection of matter (see upper-right
panels of Fig. \ref{fig:rho_2D_zoom}). After merger, a differentially
rotating object is produced whose mass is larger than the maximum mass of
rotating configuration and that is normally referred to as hypermassive
neutron star (HMNS) in the case of hadronic EOSs, but that should dubbed
hypermassive quark star (HMQS) in the case of the MIT2cfl EOS (see
lower-left and lower-right panels of Fig. \ref{fig:rho_2D_zoom}).

The dynamics of this post-merger phase is again rather similar between
the two EOSs and follows well-known behaviours in terms of dynamical
ejecta, tidal forces and shock heating (see, \eg \cite{Bovard2017,
  Radice2018a, ShibataRev19}). In particular, the dynamical ejection of
matter takes place mostly at low latitudes (\ie near the equatorial
plane, as it can be seen in the $(x,z)$ sections of Fig.
\ref{fig:rho_2D_zoom}). On the other hand, the material at high latitudes
(\ie near the polar directions, as it can be seen in the $(x,z)$ sections
of Figs.  \ref{fig:s_2D_zoom}) is considerably hotter and with a higher
entropy per baryon. Furthermore, strong shocks taking place in the HMQS
and in the HMNS lead to a pulsed dynamical mass ejection, which is
apparent in the $(x,y)$ sections of Figs. \ref{fig:rho_2D_zoom},
\ref{fig:s_2D_zoom}, where it appears in terms of a striped spiral-arm
structure. Essentially all of this matter is gravitationally unbound and
by being both hot and with high entropy, it will be involved in the
subsequent nucleosynthetic processes that cannot be modeled here.

Overall, when inspecting the full set of
Figs. \ref{fig:rho_2D_zoom}--\ref{fig:s_2D_zoom}, it is apparent that
although no major qualitative difference emerges -- as one would have
expected given the very similar bulk properties of the components in the
two binaries -- there are some quantitative differences in the dynamics
of quark-star binaries and hadron-star binaries.  Postponing a more
detailed comparison to Sec. \ref{ssec:ejecta}, we can already appreciate
a first important result of our comparative study: mass loss is smaller
in SQM binaries than in hadronic binaries. On the other hand, the
differences in temperature and entropy are much smaller and of a factor
of a few, at most.

%------------------------------------------------------------------------
%------------------------------------------------------------------------
\subsection{Gravitational-wave emission}
\label{ssec:gw}
%------------------------------------------------------------------------
%------------------------------------------------------------------------

The emission of gravitational waves from merging binaries of compact
objects is obviously one of the most interesting outcomes of these
processes and can be used to extract important information on the status
of matter when these compact objects are stars. It is therefore natural
to investigate whether the gravitational-wave signal computed in our
simulations can be exploited to distinguish the dynamics of strange
quarks stars from that of hadronic stars.

As customary, we can compute the strongest component of the
gravitational-wave signal by extracting the $\ell=m=2$ mode of the
Weyl-curvature scalar $\psi_4$ from our simulations, so that the GW
amplitudes in the corresponding mode and in the two polarization $+$ and
$\times$ can be obtained by integrating $\psi_4$ twice in time
\begin{eqnarray}
h^{22}_{+} - i h^{22}_{\times} = \int^t_{-\infty}dt^\prime \int_{-\infty}^{t^\prime} 
dt^{\prime\prime} (\psi_4)_{22}\,.
\label{eq:gw}
\end{eqnarray}

\begin{figure}[t]
  \centering
  \includegraphics[width=0.49\textwidth]{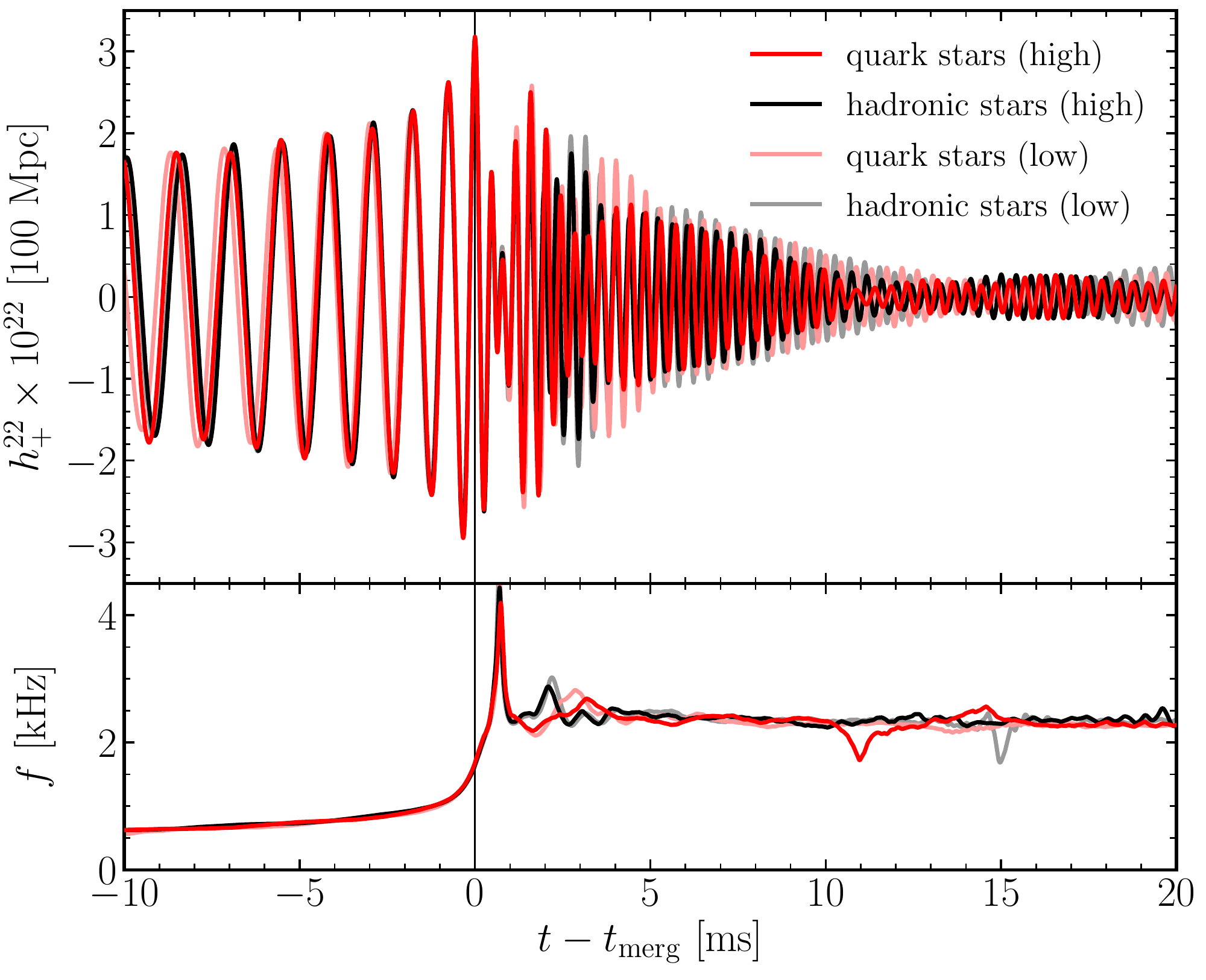}
  \caption{\textit{Top panel:} gravitational-wave strain in the
    $\ell=2=m$ mode and in the $+$ polarisation, $h^{22}_+$. Curves of
    different shading refer to different resolutions, with the
    high-resolution data being shown with the darkest shade. The vertical
    black line marks the merger time, which corresponds to the time of
    maximum amplitude, and is used to align the waveforms in
    phase. \textit{Bottom panel:} instantaneous gravitational-wave
    frequencies relative to the strain in the top panel. The same
    convention is followed for the line types.}
  \label{fig:gw_fre}
\end{figure}

\begin{figure}[t]
  \centering
  \includegraphics[width=0.49\textwidth]{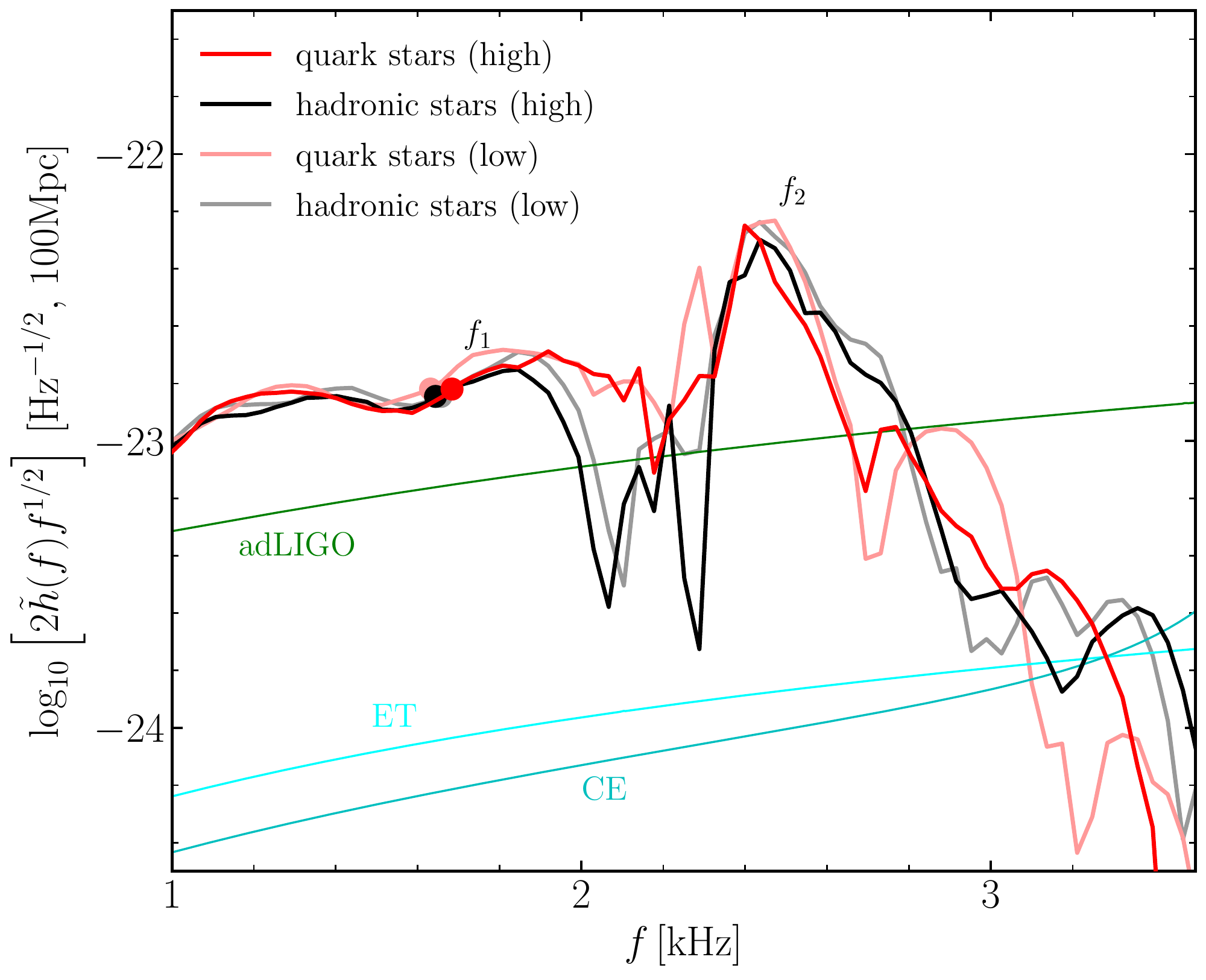}
  \caption{Effective PSDs of the quark-star (red lines) and hadron-star
    binaries (black lines) at different resolutions. Also reported with
    filled circles are the frequencies at merger $f_{\rm max}$, as well
    as the sensitivity curves of advanced and next-generation
    gravitational-wave detectors such as advanced LIGO (adLIGO), the
    Einstein Telescope (ET), or Cosmic Explorer (CE). Note the presence
    of precise spectral features, $f_1$ and $f_2$, whose frequencies are
    shown in Table~\ref{tab:gw}.}
  \label{fig:psd}
\end{figure}

Figure~\ref{fig:gw_fre}, reports in its top panel the gravitational-wave
strains $h^{22}_+$ for both the quark-star binary (red lines) and the
hadron-star binary (black). As in previous figures, different shades
refer to the two resolutions employed, with the high resolutions being
indicated with a full shading. The various waveforms are aligned at the
time of the merger, that is, the time when the amplitude reaches its
first maximum. The bottom part of Fig.~\ref{fig:gw_fre}, on the other
hand, reports, the corresponding evolution of the instantaneous
gravitational-wave frequencies using the same convention for the line
types. Overall, the gravitational-wave information provided in
Fig.~\ref{fig:gw_fre} remarks that the inspiral part of the binary
dynamics is very similar between the two types of stars, with the
frequency evolutions that are essentially the same in this phase and up
to the merger. However, small differences do develop after the merger
and, as we will comment below, they can be used to extract important
signatures on the occurrence of the merger of a quark-star binary.

\begin{figure*}[t]
  \centering
  \includegraphics[width=0.45\textwidth]{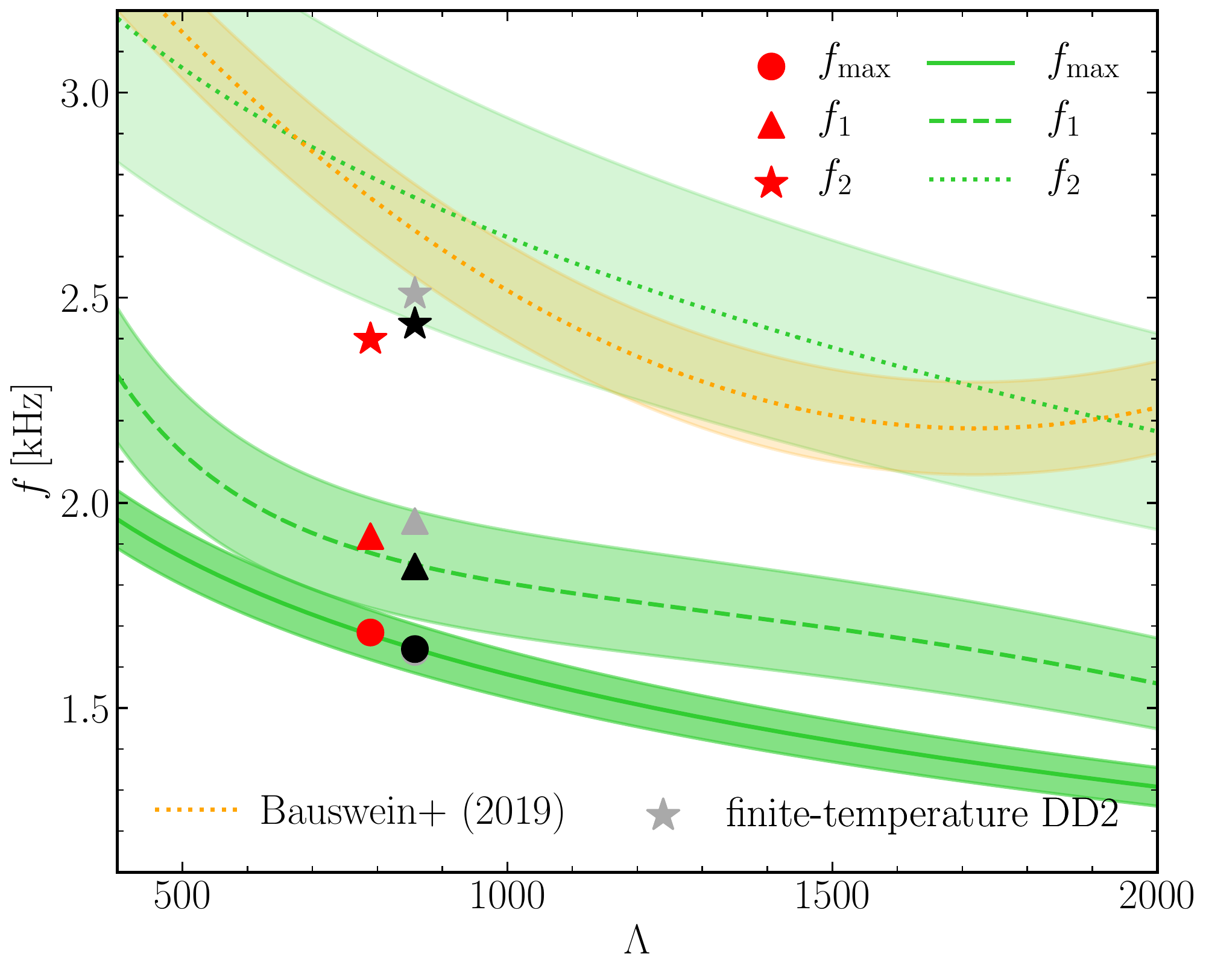}
  \hspace{0.5cm}
  \includegraphics[width=0.45\textwidth]{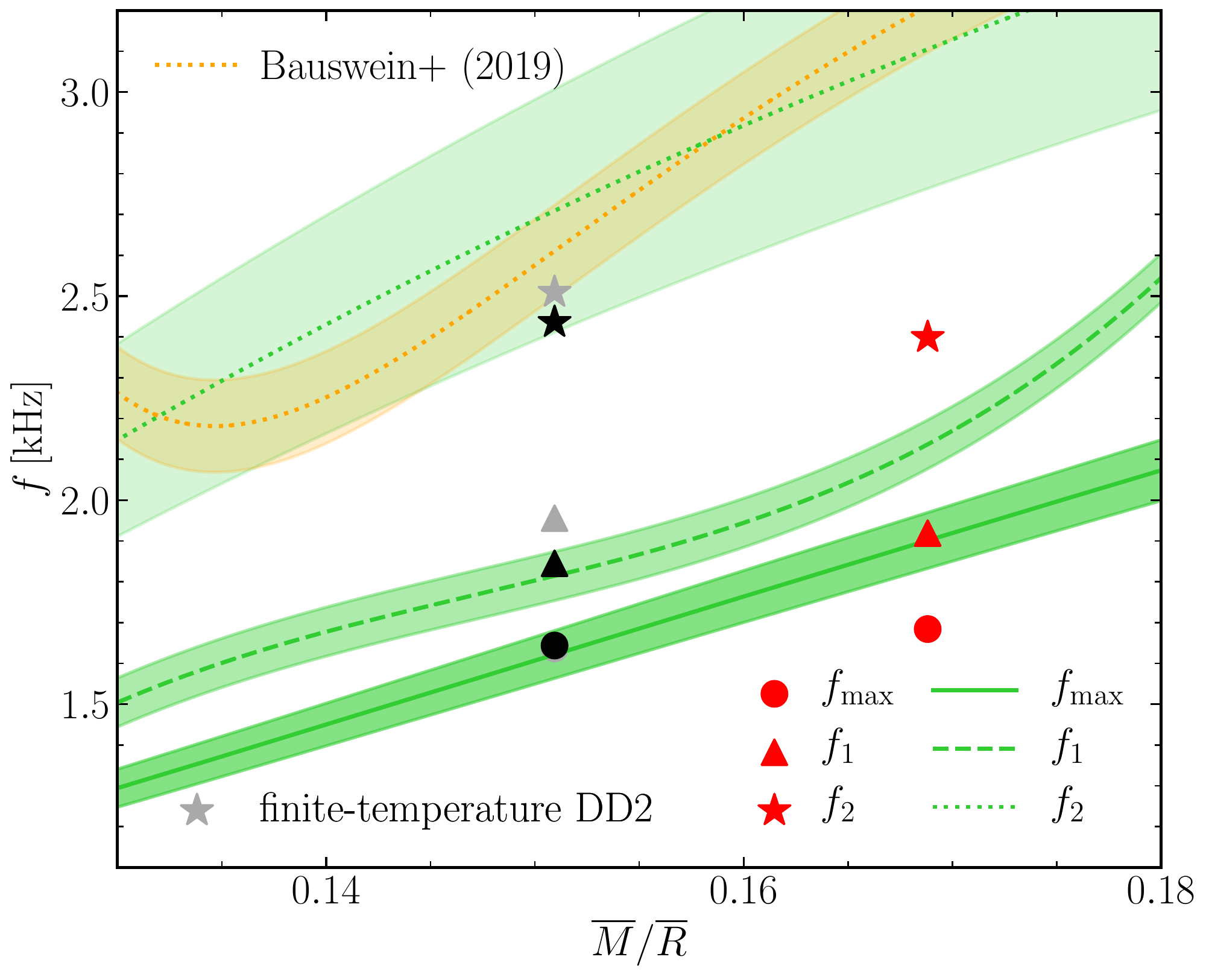}
  \caption{\textit{Left panel:} Quasi-universal relations for the
    frequency at merger $f_{\rm max}$ and the two post-merger frequencies
    $f_1$ and $f_2$ when shown as a function of the tidal deformability
    $\Lambda$ ([\cf Eqs. \eqref{eq:fmax}, \eqref{eq:f1_1}, and
      \eqref{eq:f2_1}]). Shown with symbols are the values of the
    frequencies measured from the high-resolution simulations in the case
    of quark-star binaries (red symbols) or hadron-star binaries (black
    symbols). Reported for comparison with gray symbols are the
      frequencies when considering the temperature-dependent version of
      the DD2 EOS. \textit{Right panel:} same as in the left panel but
    when the quasi-universal relations are expressed in terms of the
    average stellar compactness $\mathcal{C}$ ([\cf Eqs. \eqref{eq:f1_2}
      and \eqref{eq:f2_2}]). In both panels the shading refers to the
    uncertainties in the relations and the orange lines report the
    functional fitting in Ref. \cite{Bauswein2019}.}
  \label{fig:universal}
\end{figure*}

In order to highlight the differences that emerge after the merger we
follow the methodology presented in Ref.~\cite{Takami2015} to process the
post-merger GW signal and perform a spectral analysis whose results are
shown in Fig. \ref{fig:psd}. First of all, we collect the GW signal in
the time interval $t\in [-1500, 4000]\,M_\odot \sim [-7.39, 19.70]
\,{\rm ms}$, and perform a Fourier transformation with a symmetric
time-domain Tukey window function, with parameter $\alpha=0.25$. This
window function helps eliminating spurious oscillations of the computed
PSD. Furthermore, since we are interested only in the post-merger signal,
we employ a fifth-order high-pass Butterworth filter for the
low-frequency part of the signal with a cutoff frequency set to be
$f_{\rm cut} = f(t-t_{\rm merg} = -7.38\,{\rm ms})+0.5\,{\rm kHz}$. In
this way we compute the effective PSD as~\cite{Takami2015}
\begin{eqnarray}
\tilde{h}(f) := \sqrt{\frac{|\tilde{h}_{+}(f)|^2 +
    |\tilde{h}_{\times}(f)|^2}{2}}\,,
\label{eq:psd}
\end{eqnarray}
where $\tilde{h}_{+}(f)$ and $\tilde{h}_{\times}(f)$ are the Fourier
transforms of the to two gravitational-wave strains. At the same time, it
is possible to compute signal-to-noise ratio (SNR) as
\begin{eqnarray}
{\rm SNR} := \left[\int_0^{\infty}
  \frac{|2\tilde{h}(f)f^{1/2}|^2}{S_h(f)} \frac{df}{f}\right]^{1/2}\,,
\label{eq:snr}
\end{eqnarray}
where $S_h(f)$ is the noise PSD of a given gravitational-wave
detector. 

Figure~\ref{fig:psd} reports the effective PSDs of the two binaries (red
and black lines for the quark-star and the hadron-star binaries,
respectively) at the two different resolutions (dark and light shading
for the high and low resolutions, respectively), and the frequencies at
merger $f_{\rm max}$ (filled circles; these frequencies are also used to
match the amplitudes of the PSDs obtained at different resolutions). Also
reported in Fig. \ref{fig:psd} are the sensitivity curves of advanced and
next-generation gravitational-wave detectors such as advanced LIGO
(adLIGO), the Einstein Telescope (ET)~\cite{Punturo:2010, Punturo2010b},
or Cosmic Explorer (CE)~\cite{Abbott17}. In this way, it is possible to
appreciate the presence of precise spectral frequencies, \ie
gravitational-wave spectral lines, which are named as $f_1$ and $f_2$
after the convention in \cite{Takami2014}, and whose values are shown in
Table~\ref{tab:gw}. Also reported as filled solid circles on top of the
various PSDs are the gravitational-wave frequencies at the merger (\ie at
amplitude maximum) $f_{\rm max}$. Note that a third frequency, $f_3$ can
be obtained from the approximate relation $2f_2 \approx f_1 + f_3$, which
models the dynamics of the two stellar cores in terms and repeated
collisions and bounces while the HMNS/HMQS rotates.

As first pointed out in Ref. \cite{Read2013}, the values of $f_{\rm max}$
can be used as an important proxy of the EOS. Furthermore, this frequency
was shown to follow a quasi-universal behaviour in terms of the tidal
deformability. On the other hand, both Fig. \ref{fig:gw_fre} and
Fig. \ref{fig:psd} clearly indicate that the values of $f_{\rm max}$ are
rather similar for both quark and hadronic stars. This is because,
although the radii of the stars in the two binaries are rather different,
\ie $R_{_{\rm QS}} \simeq 11.8\,{\rm km}$ and $R_{_{\rm HS}}\simeq
13.2\,{\rm km}$ for the quark (QS) and hadron star (HS), respectively
(see Table \ref{tab:gw}), the corresponding tidal deformabilities are
very similar, \ie $\Lambda_{_{\rm QS}} \simeq 790$ and $\Lambda_{_{\rm
    HS}} \simeq 860$. Indeed, in Ref.~\cite{Rezzolla2016} and using a
large set of purely hadronic EOSs, a universal relation was found between
the $f_{\rm max}$ and tidal deformability\footnote{In
Ref. \cite{Rezzolla2016}, the original form of Eq. \eqref{eq:fmax} was
given in terms of the tidal deformability coefficient $\kappa_2^{\rm T}$,
which is trivially related to $\Lambda$ and in the case of equal-mass
binaries as $\kappa_2^{\rm T} = \tfrac{3}{16}\Lambda$.}, namely 
\begin{equation}
\log_{10}\left(\frac{f_{\rm max}}{{\rm Hz}}\right) \approx a_0 + a_1
\Lambda^{1/5} - \log_{10}\left(\frac{2\overline{M}}{M_\odot}\right)\,,
\label{eq:fmax}
\end{equation}
where $a_0 = 4.186, a_1 = -0.140$, and $\overline{M}$ represent the
average mass at infinite separation (in our simulations and for both
binaries, $\overline{M}=1.35\, M_\odot$).

The quasi-universal relation \eqref{eq:fmax} is shown with a green solid
line in the left panel of Fig. \ref{fig:universal} as a function of the
tidal deformability $\Lambda$ and in the right panel as a function of the
average stellar compactness $\mathcal{C} := \overline{M}/\overline{R}$,
where $\overline{R}$ is the average radius. We recall that this is
possible because another quasi-universal relation exists relating the
tidal deformability $\Lambda$ with the stellar compactness $\mathcal{C}$,
namely \cite{Yagi2017, Raithel2018}
\begin{eqnarray}
\mathcal{C} = 0.360 - 0.0355\,\ln\Lambda + 0.000705\,(\ln\Lambda)^2\,.
\label{eq:c-lambda}
\end{eqnarray}

Also shown in Fig.~\ref{fig:universal} with red and black symbols are the
values of the measured gravitational-wave frequencies at merger $f_{\rm
  max}$ for binary quark stars (red filled circles) and binary hadronic
stars (black filled circles), respectively. When comparing the numerical
results with the expected quasi-universal relation with the tidal
deformability (left panel of Fig. \ref{fig:universal}), it is clear that
the match is very good despite the very different nature of the two
binaries and, in particular, the large difference in radii. This result
may appear bizarre at first as one would expect that the
gravitational-wave frequency at merger is directly related to the stellar
radius and indeed the ``contact'' frequency is normally computed as the
Keplerian frequency when the two stars have a separation which is twice
the stellar radius $f_{\rm cont} := \mathcal{C}^{3/2}/(2\pi
\overline{M})$ \cite{Damour:2012, Takami2014}. However, as recently
pointed out in Ref.  \cite{Ng2020}, the gravitational-wave frequency at
merger $f_{\rm max}$ is closely related to the quadrupolar ($\ell = 2$)
$F$-mode oscillation ($F_{2F}$) of nonrotating and rotating models;
furthermore, $F_{2F}$ is can also be expressed quasi-universally in terms
of the tidal deformability, so that the results shown in the left panel
of Fig. \ref{fig:universal} are essentially stating that quark and
hadronic stars with similar tidal deformability -- and thus similar
properties of their dense cores -- have similar quadrupolar $F_{2F}$
oscillation modes and, hence, similar gravitational-wave frequency at
merger $f_{\rm max}$. This is another important result of our comparative
study: quark-star binaries have merger frequencies similar to those of
hadron-star binaries with comparable tidal deformability. In turn, this
implies that it may be difficult to distinguish the two classes of stars
based only on the gravitational-wave signal during the inspiral.

Note, however, that when expressed in terms of the stellar compactness
(see right panel of Fig. \ref{fig:universal}), the match between the
measured merger frequencies and the expectations from the quasi-universal
relations is reasonably good in the case of hadronic stars, but it is
much worse for quark-star binaries. This hints to the fact that while
expression \eqref{eq:fmax} could be used both for hadronic and quark-star
binaries, the corresponding expression in terms of the stellar
compactness needs to be corrected in the case of quark-star
binaries. We will return to this point below.

Figure ~\ref{fig:universal} also shows additional quasi-universal
relations for the $f_1$ and $f_2$ frequencies, both as a function of the
tidal deformability (left panel) and of the average stellar compactness
(right panel). For the first, low-frequency peak, the universal relations
are expressed respectively as \cite{Takami2015, Rezzolla2016} 
\begin{eqnarray}
f_1 & \approx & b_0 + b_1\, \mathcal{C} + b_2\, \mathcal{C}^2 +
b_3\mathcal{C}^3\ {\rm kHz}\, , \label{eq:f1_1} \nonumber \\ \\
f_1 & \approx & c_0 + c_1\, \Lambda^{1/5} + c_2\, \Lambda^{2/5} + c_3\,
\Lambda^{3/5}\ {\rm kHz} \,, \nonumber \\
\label{eq:f1_2}
\end{eqnarray}
where $b_0 = -35.17, b_1 = 727.99, b_2 = -4858.54, b_3 = 10989.88, c_0 =
45.19, c_1 = -31.11, c_2 = 7.50$, and $c_3 = -0.61$. Similarly, for the
second and most powerful feature in the post-merger spectrum, we use
\cite{Rezzolla2016} 
\begin{align}
\label{eq:f2_1}
f_2 & \approx 5.832 - 0.800\, \Lambda^{1/5}\ {\rm kHz}\,.& \\
\label{eq:f2_2}
f_2 & \approx 5.832 - 123.016 \exp\left({-\sqrt{56.738\, \mathcal{C}
  + 4.930}}\right)\ {\rm kHz}\,,& 
\end{align}
where for the second relation \eqref{eq:f2_2} we have used expression
\eqref{eq:c-lambda}. These frequencies are shown in
Fig. \ref{fig:universal} with green lines and shadings together with
their uncertainties are estimated in Ref. \cite{Rezzolla2016};
furthermore, they are complemented by the expression for the $f_2$
frequency proposed in Eq. (1) of Ref. \cite{Bauswein2019}, where it is
dubbed $f_{\rm peak}$ (orange line and shading).

A comparison with numerical data clearly shows that the universal
relations provide a rather accurate representation in the case of
hadronic stars, both when expressed in terms of the tidal deformability
(left panel) and of the average stellar compactness (right panel). The
match remains very good also in terms of quark stars, but only when the
universal relations are given in terms of the tidal deformability (in the
case of the $f_1$ frequencies, the relative differences are of $2.15\%$
and $0.22\%$ for binary quark stars and binary hadron stars,
respectively). Furthermore, the match of the $f_2$ frequency of the
hadron-star binary (black filled star) with the quasi-universal relation
is only marginally acceptable when using expressions \eqref{eq:f2_1} and
\eqref{eq:f2_2}, while it is below the value estimated by
Ref. \cite{Bauswein2019} (reported for comparison with gray symbols
  are the frequencies when considering the temperature-dependent version
  of the DD2 EOS).  However, the differences can be quite substantial
when the numerical data for binary quark stars is compared with
expressions \eqref{eq:f1_1}, \eqref{eq:f1_2} for the $f_1$ frequency or
with expressions \eqref{eq:f2_1}, \eqref{eq:f2_2} for the $f_2$
frequency.

We can summarise these findings into another notable result of our
comparative investigation: quark-star binaries have post-merger
frequencies that obey quasi-universal relations derived from hadron-star
binaries in terms of the tidal deformability, but not when expressed in
terms of the average stellar compactness. Hence, it may be difficult to
distinguish the two classes of stars based only on the post-merger
frequencies and if no information on the stellar radius is available. As
conjectured above, this behaviour seems to suggest that new universal
relations in terms of the stellar compactness need to be derived when
consider binaries of SQM. This conjecture can only be verified when
simulations with other EOSs for the SQM have been performed.

%------------------------------------------------------------------------
%------------------------------------------------------------------------
\subsection{Properties of the ejected material}
\label{ssec:ejecta}
%------------------------------------------------------------------------
%------------------------------------------------------------------------

As mentioned in the Introduction, besides gravitational waves, another
important product of the merger of a binary system of compact stars is
represented by the ejected matter as this plays an important role in
subsequent nucleosynthesis and in the kilonova emission. We have also
already remarked that the lack of a consistent treatment of the
evaporation process of SQM to nucleons prevents us from establishing what
is the actual role played by the ejected matter from the binary
quark-star merger. We recall that the study of
Ref.~\cite{Bucciantini2019} has concluded that only a small amount of
quark matter can survive evaporation and thus survive to yield a very low
density of SQM in the galaxy. At the same time, the use of a hybrid-EOS
approach to handle the thermal effects in the hadronic DD2 EOS -- and
that we have employed to achieve a consistent picture with the MIT2cfl
EOS -- prevents us from making considerations on the nucleosynthetic
yields and kilonova emission also for the binary hadron-star merger.

\begin{figure}[t]
  \centering
  \includegraphics[width=0.49\textwidth]{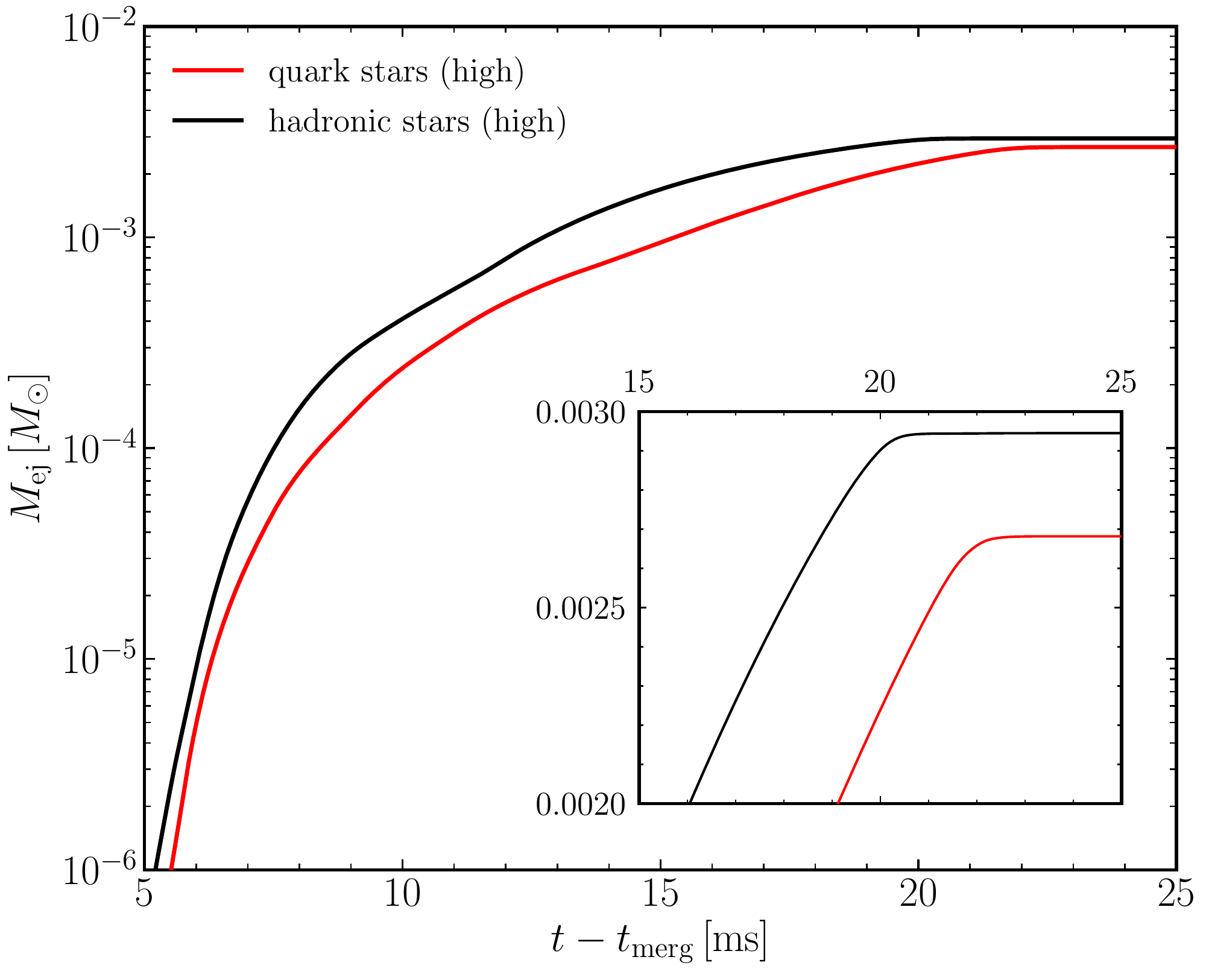}
  \caption{Ejected rest-mass as a function of time as measured by a
    detector placed at a radius of $500\, M_\odot \simeq 750\, {\rm
      km}$. Lines of different colours follow the same convention of the
    previous figures: red for quark-star binaries and black for
    hadron-star binaries. Note that the data refers to the
    high-resolution simulations and that the ejected mass for the
    quark-star binaries is smaller.}
  \label{fig:outflow}
\end{figure}

Notwithstanding these limitations, we can nevertheless obtain interesting
comparative measurements of the ejected matter in terms of bulk
properties, such as: the total amounts of ejected rest-mass for the two
classes of stars and the corresponding distributions in velocity and
entropy. To this scope, and as done routinely in this type of
simulations, we place a series of detectors in terms of spherical
coordinate 2-spheres at different distances from the origin and measure
the amount of matter that crosses such detectors. In this way,
considering gravitationally unbound the matter satisfying the so-called
``geodesic'' criterion, namely, matter whose covariant time component of
the four-velocity is $u_t< -1$ (see \cite{Bovard2016} for a detailed
discussion of various criteria for unbound matter) we can measure the
effective mass lost by the binaries. This is shown as a function of time
from the merger in Fig.~\ref{fig:outflow} for a detector placed at a
radius of $500\, M_\odot \simeq 750\, {\rm km}$, where lines of different
colours follow the same convention of the previous figures. When comparing
the results of the ejected mass from the quark-star binaries and from the
hadron-star binaries, the difference is apparent, with the former having
$M_{\rm ej} = 2.68\times 10^{-3}\,M_{\odot}$ and the latter $M_{\rm ej} = 
2.94\times 10^{-3}\,M_{\odot}$. 

\begin{figure*}[t]
  \centering
  \includegraphics[width=0.49\textwidth]{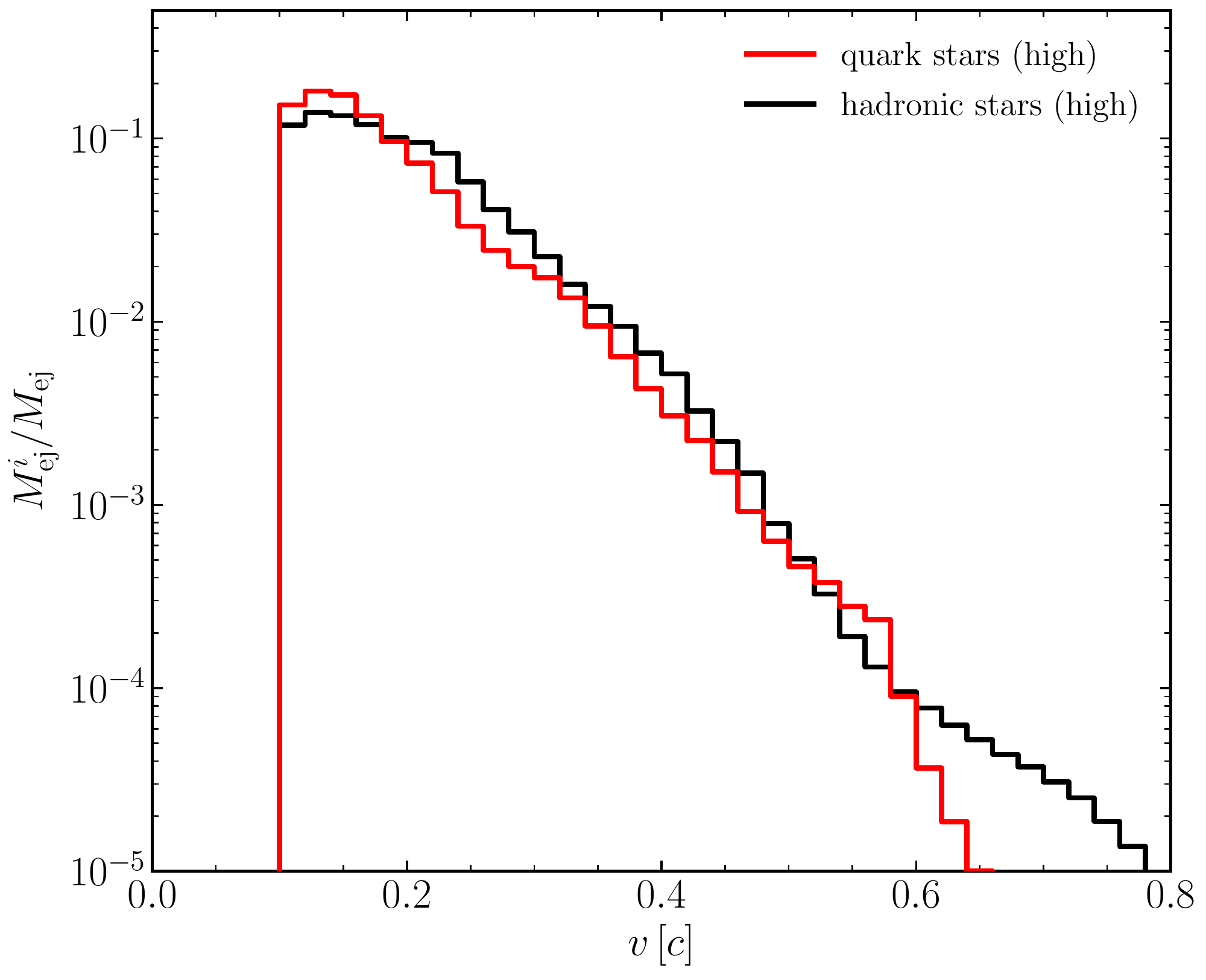}
  \includegraphics[width=0.49\textwidth]{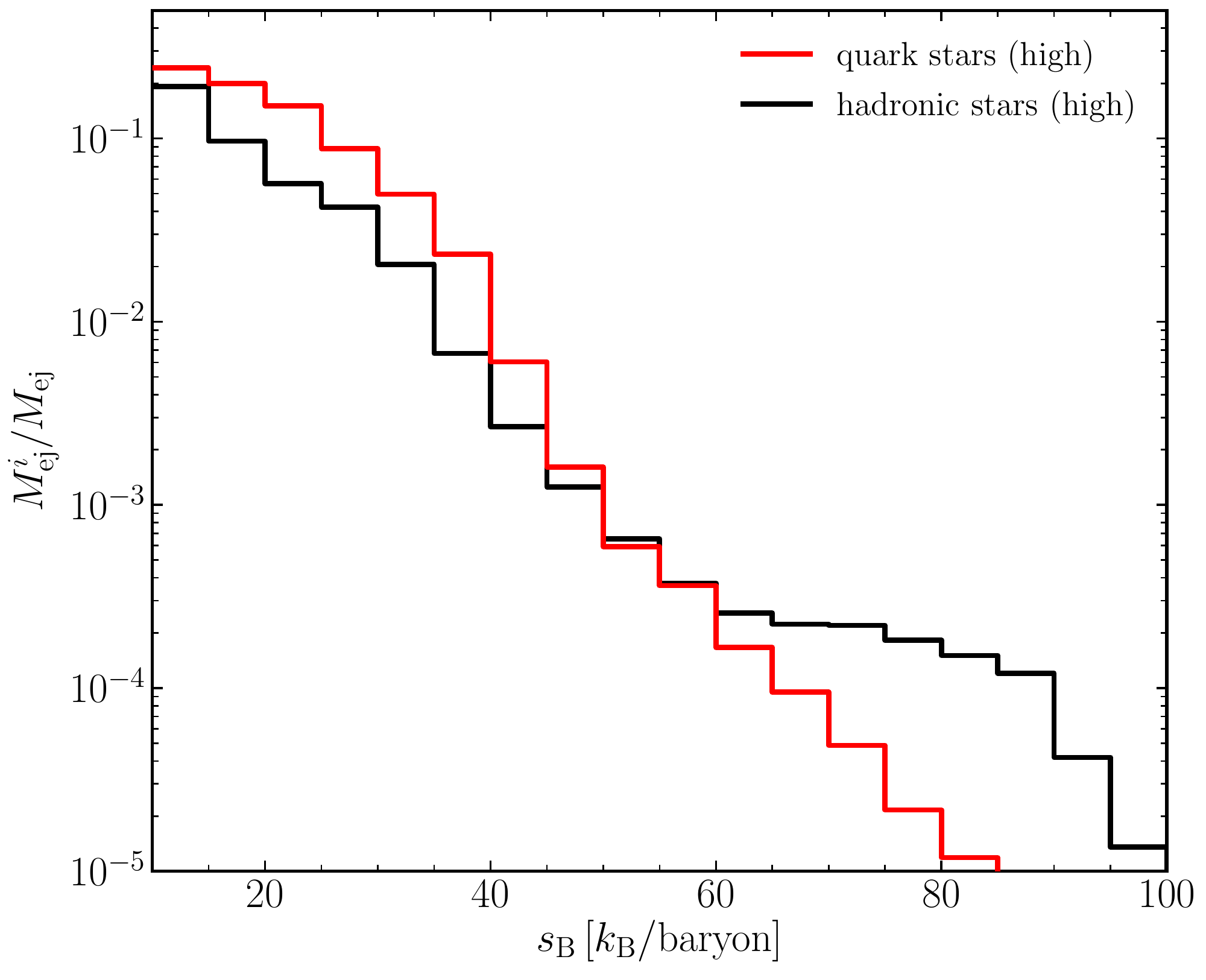}
  \caption{Distributions of the ejected matter in terms of the norm of
    the three-velocity (left panel) and of the entropy per baryon (right
    panel). Lines of different colours follow the same convention of the
    previous figures: red for quark-star binaries and black for
    hadron-star binaries. Note that the data refers to the
    high-resolution simulations and that quark-star binaries have much
    reduced tails in both quantities.}\label{fig:outflow_dist}
\end{figure*}

There are at least two explanations behind this difference. The first is
that, as mentioned repeatedly, SQM is characterised by self-bound
properties and generally harder to eject. Second, quark stars are
intrinsically more compact and hence the merged object will be subject
both to stronger dynamical shocks but also produce stronger
gravitational fields from which it is harder to eject matter. Third,
and most importantly, a rest-mass difference is present between SQM and
hadronic matter. We recall, in fact, that the average rest-mass per
baryon of SQM we adopt here $850\,{\rm MeV}$, which is smaller than the
corresponding value for hadronic matter, $940\,{\rm MeV}$.

Proceeding further in our comparison, we report in
Fig.~\ref{fig:outflow_dist}, two distributions of the ejected matter in
terms of the norm of the three-velocity (left panel) and in terms of
entropy per baryon (right panel). Note that both distributions are
relative to the ejected matter and hence provide a measure of the
fraction of the ejecta with given properties. Concentrating first on the
left panel, it is clear that the two distributions are very similar and
that differences appear only in the high-velocity tails, \ie $v\gtrsim
0.6$, which are larger by almost one order of magnitude in the case of
binary hadron-star mergers. Although the actual amount of matter in these
tails is tiny (\ie $\lesssim 10^{-8}\,M_{\odot}$), they could play a role
when interacting with the interstellar medium and produce a
re-brightening of the afterglow signal \cite{Hotokezaka2018} (see
\cite{Most2020e} for a discussion about the role of fast ejecta).

When considering instead the distribution in entropy (right panel of
Fig.~\ref{fig:outflow_dist}) -- and bearing in mind that the these
measurements are the same for the two classes of stars but equally
approximate -- it is possible to appreciate that quark-star mergers
overall produce ejected matter with larger entropy. At the same time,
hadron-star mergers are able to eject matter with very high entropy, \ie
$s_{_{\rm B}} \gtrsim 60\,k_{_{\rm B}}/{\rm baryon}$, and about one order
of magnitude more than quark-star mergers. These differences on the
entropy distributions may have a potential effects on the subsequent
kilonovae observation, but the degree to which this will happen remains
unclear until the quark-evaporation mechanism is properly taken into
account, more sophisticated EOSs are devised to estimate the impact of
SQM in binary mergers\footnote{We note that the MIT2cfl EOS considered
here and introduced in \cite{Zhou2017} does not include an electron
fraction, making it thus impossible to account for the evolution of the
electron fraction in the ejected matter.}, and more systematical
investigations are carried out.

%------------------------------------------------------------------------
%------------------------------------------------------------------------
\section{Conclusion}
\label{sec:conclusion}
%------------------------------------------------------------------------
%------------------------------------------------------------------------

Since the hypothesis that SQM is the ground state of matter has been
formulated more than four decades ago, a vast literature has been
developed to investigate this scenario and consider its viability
against the astronomical observations. In this way, a very large number
of works have been presented in which the possibility that SQM could lead
to compact stars, \ie quark stars, has been explored in the greatest
detail. Of this vast literature, however, only a very small fraction has
been dedicated to the study of the dynamics of binary systems of quark
stars. The scarcity of studies of this scenario, which is particularly
striking after the detection of gravitational waves and of
electromagnetic counterparts from the merger of low-mass compact binaries
such as GW170817, probably has a double origin. From an observational
side, there is the expectation that a merger of binary quark stars cannot
be responsible of the observed kilonova emission in GW170817 and of the
subsequent nucleosynthesis. From a theoretical side, on the other hand,
the modelling of the inspiral and merger of a binary of quark stars is
particularly challenging as it is difficult to properly capture the large
discontinuity that appears at the surface of these self-bound stars.

The first of these difficulties has been recently addressed in a number
of recent works that have invoked a quark-evaporation process into
hadrons that could have taken place as a result of the high temperatures
reached after the merger. In this case, therefore, most of ejected SQM
from the quark-star binary would have evaporated into nucleons and
therefore could have contributed to the kilonova signal in AT2017gfo
\cite{Bucciantini2019, DePietri2019, Horvath2019}. Addressing the second
difficulty, on the other hand, is the purpose of this work, where we
  have presented a suitable definition of the baryonic mass of SQM and a
  novel technique in which a very thin crust has been added to the EOS to
  produce a smooth and gradual change of the specific enthalpy across the
  crust. The new approach also allows to use different values for the
  baryonic mass, which can be introduced via a suitable rescaling of the
  hydrodynamical variables that are evolved.

The introduction of this crust, whose rest-mass content is minute, \ie
$\sim 5\times 10^{-3}\,M_\odot$, and whose spatial extension is
restricted to two grid cells only, has been carefully tested by
considering the oscillation properties of isolated quark stars. In this
way, it was possible to show that the dynamical and simulated response of
the quark stars matches accurately the perturbative predictions and that
the match becomes increasingly accurate as the numerical resolution is
increased. This validation, which has been introduced numerous times in
the past when considering neutron stars, has levels of accuracy that are
comparable with those obtained with hadronic stars.

Using this novel technique we have been able to carry out the first fully
general-relativistic simulations of the merger of binary strange quark
stars. In addition, in order to best assess the feature of this merging
process that are specific to quark stars, we have carried out a
systematic comparison of the dynamics of quark-star binaries with the
corresponding behaviour exhibited by a binary of hadron stars having the
same mass and very similar tidal deformability, namely, a binary
described by the DD2 EOS. In this way, it was possible to highlight
several important differences between the SQM and the hadronic stars,
which represent the main results of our work. First, the dynamical
mass lost is $\sim 20\%$ smaller than that coming from a corresponding
hadronic binary.  Second, quark-star binaries have merger frequencies
similar to those of hadron-star binaries with comparable tidal
deformability. Hence, it may be difficult to distinguish the two classes
of stars based only on the gravitational-wave signal during the inspiral.
Third, quark-star binaries have post-merger frequencies that obey
quasi-universal relations derived from hadron-star binaries in terms of
the tidal deformability, but not when expressed in terms of the average
stellar compactness. Hence, it may be difficult to distinguish the two
classes of stars based only on the post-merger frequencies and if no
information on the stellar radius is available. Fourth, the matter
ejected from quark-star binaries has much smaller tails in the velocity
distributions, \ie for $v \gtrsim 0.6$; this lack of fast ejecta may be
revealed by the corresponding lack of a radio re-brightening when the
fast ejecta interact with the interstellar medium. Finally, while
quark-star binaries eject material that, on average, has larger entropy
per baryon, it also lacks the important tail of very high-entropy
material. Determining whether these differences in the ejected will able
to leave an imprint in the electromagnetic counterpart and
nucleosynthetic yields remains unclear.

The results presented here had to rely necessarily on a number of
simplifying assumptions and can therefore be improved in a number of
ways. First, by adopting EOSs for the SQM that have a proper treatment of
the temperature dependence and a nonzero electron fraction. Doing so will
allow us not only to accurately and self-consistently describe the
thermodynamical evolution after the merger, but also to determine whether
the ejected material will lead to the observed chemical
abundances. Second, by employing a larger class of EOSs so that it will
be possible to establish whether the frequency at merger and the
post-merger frequencies obey new universal relations when expressed in
terms of the stellar compactness. Finally, and most importantly, by
adopting a detailed and quantitative description of the quark-evaporation
mechanism, so that a consistent assessment can be made of the viability
of binary quark-star mergers as sources to the electromagnetic
counterpart in AT2017gfo. All of these improvements will be explored and
presented in future works.

\begin{acknowledgments}
  It is a pleasure to thank M. Hanauske, J. Papenfort, L.  Weih,
  A. Drago, G. Pagliara, and A. Bauswein for useful input and
  comments. Z. Zhu acknowledges support from the China Scholarship
  Council (CSC). Support also comes in part from ``PHAROS'', COST Action
  CA16214; LOEWE-Program in HIC for FAIR; the ERC Synergy Grant
  ``BlackHoleCam: Imaging the Event Horizon of Black Holes'' (Grant
  No. 610058). The simulations were performed on the SuperMUC and
  SuperMUC-NG clusters at the LRZ in Garching, on the LOEWE cluster in
  CSC in Frankfurt, and on the HazelHen cluster at the HLRS in Stuttgart.
\end{acknowledgments}

\bibliographystyle{apsrev4-2}
\bibliography{aeireferences}

\newpage
\phantom{x}
\newpage
\appendix
%------------------------------------------------------------------------
%------------------------------------------------------------------------

%------------------------------------------------------------------------
%------------------------------------------------------------------------
\section{On the value of the baryon mass}
\label{sec:baryon mass}
%------------------------------------------------------------------------
%------------------------------------------------------------------------

\begin{figure*}
  \centering
  \includegraphics[width=0.45\textwidth]{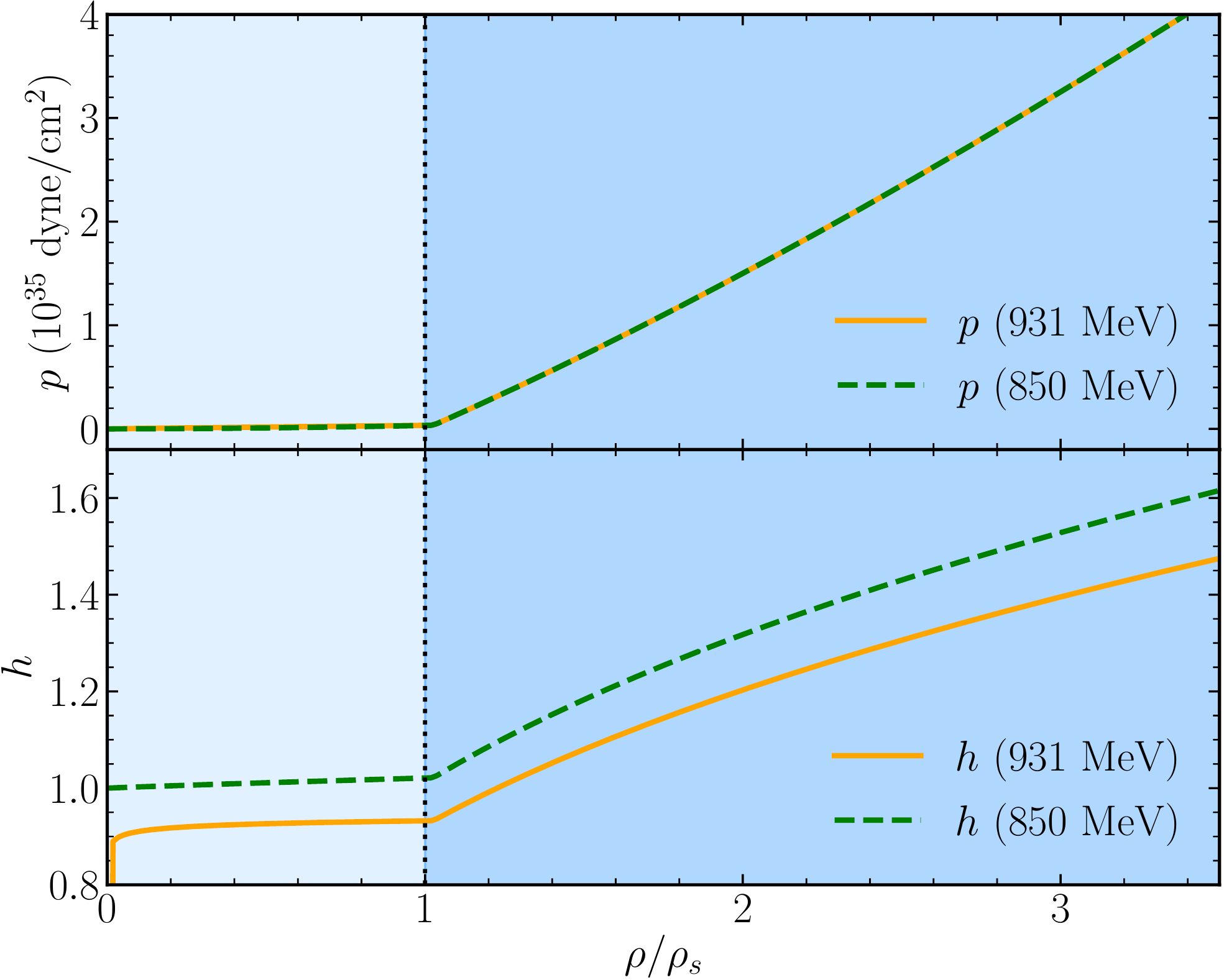}
  \hskip 0.5cm
  \includegraphics[width=0.45\textwidth]{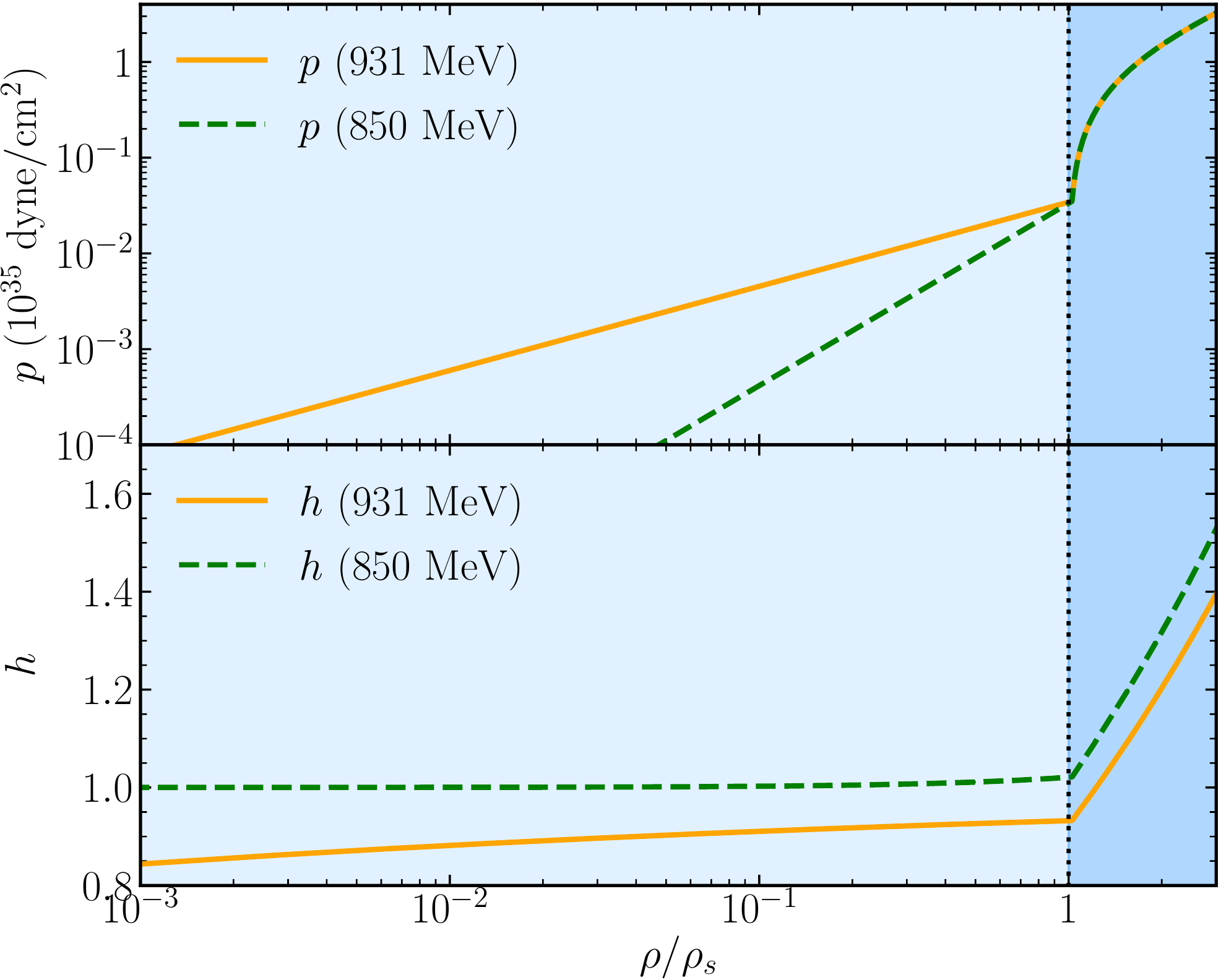}
  \caption{\textit{Left panels:} behaviour of the pressure $p$ (top part)
    and of the specific enthalpy $h$ (bottom part) of the MIT2cfl EOS
    shown as a function of the rest-mass density normalised to the
    nominal (\ie without a crust) value at the surface $\rho/\rho_s$. The
    values presented refer to the original pressure (orange solid lines)
    and to the rescaled one [\cf Eq. \eqref{eq:rescaling}] (green dashed
    lines). Note the appearance of a small jump in $h$ for very small
    densities. \textit{Right panels:} same in the right panels, but when
    the rescaled rest-mass density is shown in a logarithmic scale to
    highlight the smoother transition.}
  \label{fig:eos_eff}
\end{figure*}

This appendix is dedicated to some considerations about the value
  assumed for the baryon mass and what needs to be taken into account if
  different choices are made. We start this discussion by recalling that
  the introduction of thin crust modelled with a polytropic EOS requires
  a certain amount of care as it is necessary to ensure that proper
  physical conditions are present at the interface between the crust
  and the SQM. This interface will behave as a contact discontinuity,
  with jump condition across it given by the continuity of pressure and
  specific enthalpy, \ie ~\cite{Rezzolla_book:2013}
\begin{eqnarray}
\llbracket p\rrbracket = \llbracket h\rrbracket = 0\,, \label{eq:cdiscon}
\end{eqnarray}
where $\llbracket \phi \rrbracket$ measures the jump of the scalar
function $\phi$ across the contact discontinuity. Clearly, if $\llbracket
h \rrbracket \neq 0$, a pressure jump will develop, which will blow away
the surface layers of the quark star into the atmosphere\footnote{ We
recall that numerical-relativity codes with an Eulerian description of
the fluids require the presence of a very low-density atmosphere outside
of the compact stars for the solution of the equations of relativistic
hydrodynamics (see \cite{Rezzolla_book:2013} for a discussion). In our
setup, the atmosphere is set to be at a threshold density that is $12$
orders of magnitude below the maximum density in the star. Note that, in
contrast, Lagrangian codes do not have this requirement and can
virtually simulate regions of matter vacuum \cite{Rosswog2020}.}.

Clearly, if such a jump in the specific enthalpy is not removed, the
  quark star will rapidly diffuse away into the atmosphere, thus
  preventing a numerical evolution. However, since the specific enthalpy
  $h$ depends on the value of baryonic mass $m_{_{\rm B}}$, a potential
  jump in $h$ at the stellar surface can be removed by a suitable
  rescaling of the baryon mass. In particular, if we were interested in
  adopting the value of the baryon mass normally adopted to describe
  hadronic matter, namely, $m_{_{\rm B}}=940\,{\rm MeV}$ -- which is also
  sometimes used to mark the mass of strange quark matter (see, \eg
  \cite{DePietri2019}) -- then we would have to face the appearance of a
  small but nonzero jump in $h$ at the stellar surface.

This is shown in Fig. \ref{fig:eos_eff}, which reports with a solid
orange line the behaviour of the pressure (top panels) and of the
specific enthalpy (lower panels) as a function of the rest-mass density
normalized to the putative rest-mass density at the surface of the star
(\ie before the addition of a crust), $\rho_s$, when assuming a
  baryon mass $\tilde{m}_{_{\rm B}}=940\,{\rm MeV}$. The left panels, in
particular, adopt a linear scale in $\rho/\rho_s$, while the right panels
a logarithmic scale to highlight different properties of the EOS. Note
from the bottom-left panel that a jump in $h$ is present near
$\rho/\rho_s \simeq 0$ and a small part of the star even has a unphysical
value of the enthalpy, \ie $h<1$.

While these results already suggest that such value of the baryon
  mass should not be considered, a simple rescaling of all of relevant
  hydrodynamical quantities allows one in principle also to employ
  $\tilde{m}_{_{\rm B}}=940\,{\rm MeV}$.
In practice, defining as ``core'' whatever is the genuine interior of the
quark star, it is possible to operate a rescaling within the core
of the rest-mass density, of the specific enthalpy, and of the specific
internal energy $\epsilon$ of the type
\begin{align}
  \label{eq:rescaling}
  &\rho \to \rho_{\rm eff}\,,&
  &h \to h_{\rm eff}\,,&
  &\epsilon  \to \epsilon_{\rm eff}\,,&
\end{align}
where
\begin{align}
\rho_{\rm eff} &:= \chi \rho = \chi m_{_{\rm B}} n_{_{\rm B}}\,,&  \label{eq:effrho} \\
h_{\rm eff} &:= \frac{h}{\chi}\,,& \\
\epsilon_{\rm eff} &:= \frac{1+\epsilon}{\chi}
- 1\,, &
\label{eq:effvar_core}
\end{align}
while the energy density and pressure are left unchanged, \ie $e_{\rm
  eff} = e$, $p_{\rm eff} = p$. In essence, through this rescaling
  we effectively introduce an effective and larger baryon mass
  $m_{_{\rm B}}^{\rm eff} := \chi \tilde{m}_{_{\rm B}} = 940\,{\rm MeV}$,
  where $\chi = 940/850 \simeq 1.11$ such that $h_{\rm eff}=1$ and
  $\llbracket h_{\rm eff}\rrbracket = 0$ at the stellar surface.

The rescaling in Eqs. \eqref{eq:effrho}--\eqref{eq:effvar_core} described 
above is done only at the level of the numerical solution of the GRHD 
equations and to remove the difficulties introduced by the sharp jump at 
the stellar surface. However, after the completion of each timestep, all 
the physical quantities entering, for instance, on the right-hand-side 
of the Einstein equations, are evaluated with the physically correct value 
of $m_{_{\rm B}}$ and hence yield the physically contributions to the
energy-momentum tensor.

Combining Eqs.~(\ref{eq:effvar_core}) with the thermal part of EOS, one
can easily derive the effective quantities of cold part
\begin{align}
p_{\rm eff, c} & =  p_{\rm c}\,,&\\
h_{\rm eff, c} & =  \frac{h_{\rm c}}{\chi}\,,\\ 
\epsilon_{\rm eff, c} & = \frac{1+\epsilon_{\rm c}}{\chi} - 1\,.&
\label{eq:effcold}
\end{align}

The treatment of the EOS within the crust, on the other hand, remains the
same as before, namely, with a single polytrope joining from the stellar
surface down to the values of the density associated with the atmosphere
(see top-right panel).

The rescaling procedure described above has a simple and direct impact on
the equations of relativistic hydrodynamics that we solve
numerically. Adopting a flux-conservative formulation of the GRHD
equations, the conservation of rest-mass, energy, and momentum can be
cast into an ``effective'' Valencia formulation
\begin{eqnarray}
\partial_t (\sqrt{\gamma}\boldsymbol{U}_{\rm eff}) & = & \partial_i(\sqrt{\gamma}
\boldsymbol{F}^i_{\rm eff}) = \boldsymbol{S}_{\rm eff}\,,
\label{eq:valecia}
\end{eqnarray}
where we have employed the rescaling in
Eqs. \eqref{eq:effrho}--\eqref{eq:effvar_core}
\begin{eqnarray}
  \boldsymbol{U}_{\rm eff} & := & \begin{pmatrix} D \\
    S_j \\
    E \end{pmatrix}
  = \begin{pmatrix}               \rho_{\rm eff} W \\
    \rho_{\rm eff} h_{\rm eff} W^2 v_j \\
    \rho_{\rm eff} h_{\rm eff} W^2 - p_{\rm eff}
  \end{pmatrix}\,, \\
  \nonumber 
  \label{eq:valecia1}
\end{eqnarray}
\begin{eqnarray}
  \boldsymbol{F^i}_{\rm eff} & := & \begin{pmatrix} 
    \alpha v^i D - \beta^i D \\
    \alpha S^i_{\ j} - \beta^i S_j \\
    \alpha S^i - E\beta^i \end{pmatrix}\,,
  \\ \nonumber
  \label{eq:valecia2}
\end{eqnarray}
\begin{eqnarray}
  \boldsymbol{S}_{\rm eff} & := & \begin{pmatrix} 
    0 \\
    \alpha S^{ik}\partial_j \gamma_{ik}/2 + S_i 
    \partial_j \beta^i - E\partial_j \alpha \\
    \alpha S^{ij} K_{ij} - S^j \partial_j \alpha 
  \end{pmatrix}\,,
  \label{eq:valecia3}
\end{eqnarray}
and where $S^{ij} := \rho_{\rm eff}h_{\rm eff}W^2 v^i v^j + p_{\rm eff}
\gamma^{ij}$. Note that the quantities $\gamma_{ij}$, $\gamma$,
$\alpha$, $\beta^i$, and $K^{ij}$ are, respectively, the spatial
three-metric, its determinant, the lapse function, the shift vector, and
the extrinsic curvature.

Note that $\gamma_{ij}$, $\alpha$, $\beta^i$, and $K^{ij}$ are related to
the spacetime metric, which is computed from the Einstein equations via
the energy-momentum tensor that is always computed from un-rescaled
quantities; hence, these spacetime variables are unaffected by the
rescaling operated in the EOS. Similarly, the quantities $S^i$, $E$, and
$S^{ij}$ all depend on the product $\rho_{\rm eff} h_{\rm eff}$, where
the scaling cancels out (\ie $\rho_{\rm eff} h_{\rm eff} = \rho h$), or
on the pressure that is not the rescaled in the core (\ie $p_{\rm eff} =
p$); hence they are not modified by the rescaling. Similarly, the
three-velocity $v^i$ and the corresponding Lorentz factor $W:= (1-v^i
v_i)^{-1/2}$ are unaffected by the rescaling in the EOS. Consequently,
the whole set of GRHD equations~(\ref{eq:valecia})--(\ref{eq:valecia3})
remains unchanged in the core under the rescaling. The only exception is
represented by the first equation of the set, \ie the one involving the
conservation of rest-mass. However, since the only change is in the
effective baryon mass -- which appears in the conservation equation just
as a multiplicative constant -- the change is practically trivial.

A couple of remarks before concluding this appendix.  First, the
rescaling in Eqs. \eqref{eq:effrho}--\eqref{eq:effvar_core} described
above is done only at the level of the numerical solution of the GRHD
equations and to remove the difficulties introduced by the sharp jump at
the stellar surface. However, after the completion of each timestep, all
the physical quantities entering, for instance, on the right-hand-side of
the Einstein equations, are evaluated with the physically correct value
of $m_{_{\rm B}}$ and hence yield the physically contributions to the
energy-momentum tensor. Second, while $\chi$ is essentially arbitrary
and just depends on the value chosen for $\tilde{m}_{_{\rm B}}$, it
is essential that condition of contact discontinuity
Eq.~(\ref{eq:cdiscon}) is fulfilled at the stellar surface and that the
specific enthalpy at the stellar surface approaches unity.

\begin{figure}[t]
  \centering
  \includegraphics[width=0.90\columnwidth]{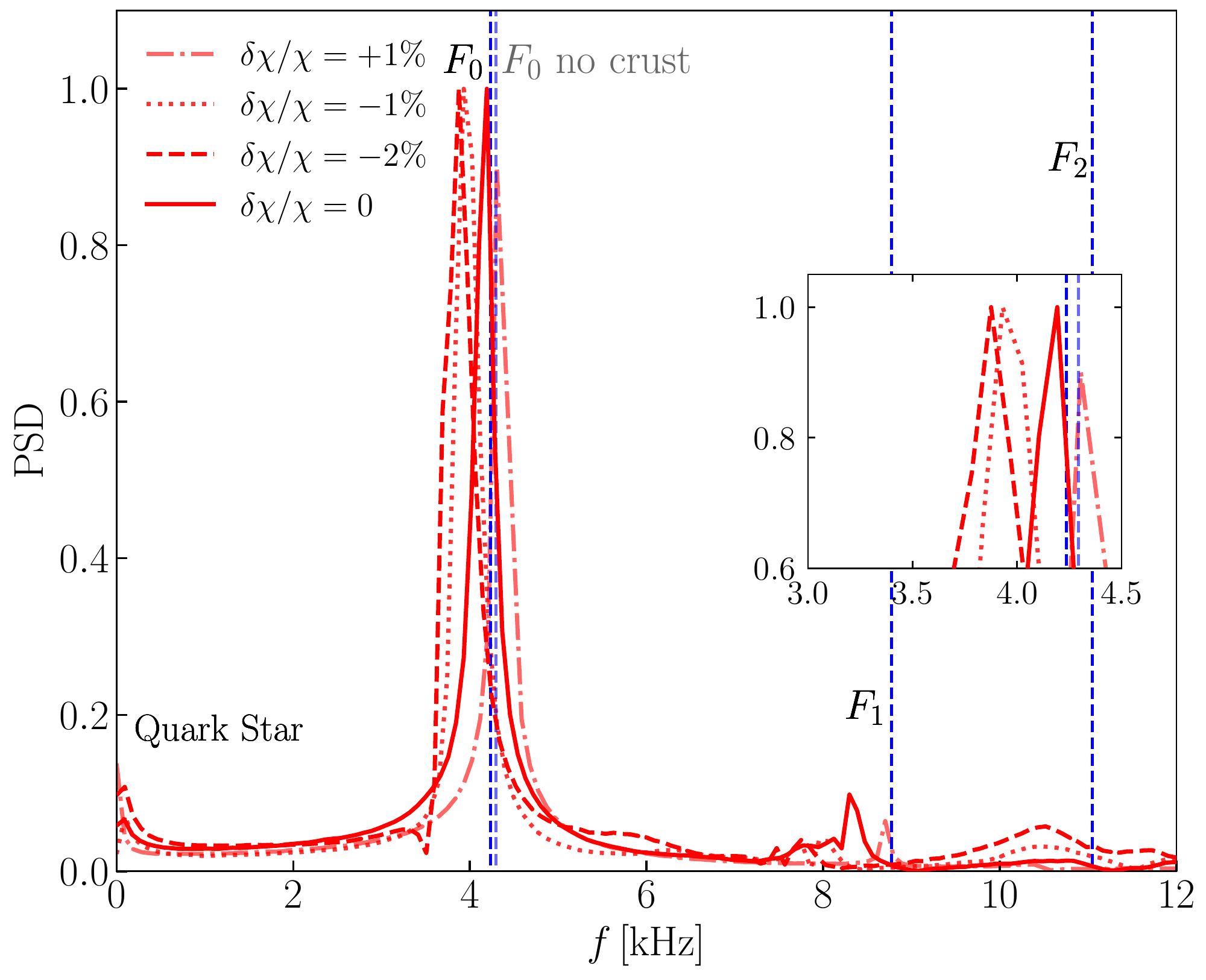}
  \caption{Same as Fig. \ref{fig:psd_oscil}, but when considering values
    of the scaling parameter $\chi$ differing by a few percent from the
    reference value.}
    \label{fig:psd_chi}
\end{figure}

In order to explore the level of flexibility in varying the value of
  the scaling constant $\chi$, we have performed simulations of isolated
  strange stars in which we have varied $\chi$ by a few percent with
  respect to its reference value. The results of this study are
  summarised in Fig.~\ref{fig:psd_chi}, which is similar to the left
  panel of Fig.~\ref{fig:psd_oscil}, but reports the PSD when increasing
  $\chi$ by $1\%$ (\ie $\delta\chi/\chi=1\%$, red dot-dashed line) or
  decreasing it by $1\%$ ($\delta\chi/\chi=-1\%$, red dotted line), while
  performing the simulation at the highest resolution. A rapid inspection
  of the figure reveals that smaller (larger) value of the scaling
  constant yield eigenfrequencies that are smaller (larger) than the
  expected value (red solid line) and that a difference in $\chi$ of only
  $1\%$ is sufficient to produce differences in the eigenfrequencies of
  $6\%$. Furthermore, when decreasing $\chi$ further (\eg
  $\delta\chi/\chi=-2\%$ as in the red dashed line) the differences with
  the perturbative results become even more severe and when
  $\delta\chi/\chi=2\%$ the specific enthalpy is below unity at the
  surface, preventing a numerical evolution. Finally, we note that
  because varying the value of $\chi$ effectively changes the slope of
  the polytropic EOS employed in the crust, and hence its thickness,
  increasing $\chi$ effectively corresponds to making the crust thinner,
  thus explaining why the evolution with slightly smaller value of $\chi$
  leads to an eigenfrequency that approaches the value of $F_0$ for a
  strange star without crust.

\end{document}